%% file: Continuous_Sample_Size_Manuscript.tex
\documentclass[11pt]{article}

\usepackage{amsmath, amsthm, amsfonts, amssymb, commath, enumerate, graphicx, bm, float, bbm, booktabs, authblk, subcaption}

\usepackage[numbers,super]{natbib}

\usepackage[inline]{enumitem}
\usepackage{setspace}

\usepackage[usenames,dvipsnames]{xcolor}

\usepackage{tikz}
\usetikzlibrary{positioning, arrows.meta, decorations.markings,calc}

\usepackage[margin=1in]{geometry}
\usepackage{hyperref}
\usepackage[noabbrev,poorman]{cleveref}
\hypersetup{breaklinks,colorlinks,allcolors=blue,pdfpagemode=UseNone}

\newenvironment{acks}[1]%
{\subsection*{\normalsize\bfseries Acknowledgements}\begin{refsize}\noindent #1}{\end{refsize}}

\makeatletter
\@ifpackageloaded{subcaption}{%
	\renewcommand*\subcaption@@label[2]{%
		\@bsphack\begingroup
		\subcaption@ORI@label#1{#2}%
		\let\SK@\@gobbletwo
		\protected@edef\@currentlabel{\csname thesub\@captype\endcsname}%
		\protected@edef\cref@currentlabel{%
			[subs\@captype][\arabic{sub\@captype}][\cref@result]%
			\csname thesub\@captype\endcsname}%
		\subcaption@ORI@label#1{sub@#2}%
		\endgroup\@esphack}%
}
\makeatother

\newlist{assumec}{enumerate}{2}
\setlist[assumec,1]{label*=\textup{I\arabic*},leftmargin=*,resume=assumec}
\setlist[assumec,2]{label=\textup{(\alph*)},leftmargin=*,ref=\textup{I\arabic{assumeci}(\alph{assumecii})}}
\crefname{assumeci}{identifiability assumption}{identifiability assumptions}
\crefalias{assumecii}{assumeci}

\newlist{assumes}{enumerate}{2}
\setlist[assumes,1]{label*=\textup{S\arabic*},leftmargin=*,resume=assumes}
\setlist[assumes,2]{label=\textup{(\alph*)},leftmargin=*,ref=\textup{S\arabic{assumesi}(\alph{assumesii})}}
\crefname{assumesi}{working assumption}{working assumptions}
\crefalias{assumesii}{assumesi}

\crefname{design}{design}{designs}
\Crefname{design}{Design}{Designs}
\crefname{model}{model}{models}
\Crefname{model}{Model}{Models}
\crefname{formula}{formula}{formulae}
\Crefname{formula}{Formula}{Formulae}
\creflabelformat{model}{(#2#1#3)}
\creflabelformat{formula}{(#2#1#3)}

\DeclareMathOperator*{\var}{Var}

\DeclareMathOperator*{\E}{E}
\newcommand{\numbereqn}{\addtocounter{equation}{1}\tag{\theequation}}
\crefname{equation}{equation}{equations}
\newcommand{\ind}[1]{\mathbbm{1}_{\left\{ #1 \right\}}}

\title{Sample size considerations for comparing dynamic treatment regimens in a sequential multiple-assignment randomized trial with a continuous longitudinal outcome}

\author[1]{Nicholas J. Seewald}
\author[2]{Kelley M. Kidwell}
\author[3]{Inbal Nahum-Shani}
\author[4]{Tianshuang Wu}
\author[5]{James R. McKay}
\author[1,3]{Daniel Almirall}

\affil[1]{Department of Statistics, University of Michigan, Ann Arbor, Michigan, USA}
\affil[2]{Department of Biostatistics, University of Michigan, Ann Arbor, Michigan, USA }
\affil[3]{Institute for Social Research, University of Michigan, Ann Arbor, MI, USA}
\affil[4]{AbbVie, North Chicago, Illinois, USA}
\affil[5]{Department of Psychiatry, Perelman School of Medicine, University of Pennsylvania, Philadelphia, Pennsylvania, USA}

\begin{document}

	\maketitle

	\begin{abstract}
		Clinicians and researchers alike are increasingly interested in how best to personalize interventions. A dynamic treatment regimen (DTR) is a sequence of pre-specified decision rules which can be used to guide the delivery of a sequence of treatments or interventions that are tailored to the changing needs of the individual. The sequential multiple-assignment randomized trial (SMART) is a research tool which allows for the construction of effective DTRs. We derive easy-to-use formulae for computing the total sample size for three common two-stage SMART designs, in which the primary aim is to compare two embedded DTRs using a continuous repeated-measures outcome collected over the entire study. We show that the sample size formula for a SMART can be written as the product of the sample size formula for a standard two-arm randomized trial, a deflation factor that accounts for the increased statistical efficiency resulting from a repeated-measures analysis, and an inflation factor that accounts for the design of a SMART.  The SMART design inflation factor is typically a function of the anticipated probability of response to first-stage treatment. We review modeling and estimation for DTR effect analyses using a repeated-measures outcome from a SMART, as well as the estimation of standard errors. We also present estimators for the repeated-measures covariance matrix for a variety of common working correlation structures.  Methods are motivated using the ENGAGE study, a SMART aimed at developing a DTR for increasing motivation to attend treatments among alcohol- and cocaine-dependent patients.	
	\end{abstract}
    
	%\doublespacing

	\section{Introduction}
	\label{sec:intro}
	Dynamic treatment regimens (DTRs) are sequences of pre-specified decision rules leading to courses of treatment which adapt to a patient's changing needs.~\cite{Kosorok2016} DTRs operationalize clinical decision-making by recommending particular treatments or intervention components to certain subsets of patients at specific times.~\cite{Chakraborty2013} Consider the following example DTR which was designed to increase engagement with an intensive outpatient rehabilitation program (IOP) for patients with alcohol and/or cocaine dependence: ``Within a week of the participant becoming non-engaged in the IOP, provide a phone-based session focusing on helping the patient re-engage in the IOP. At week 8, look back at the participant's engagement pattern over the past eight weeks. If the participant continued to not engage, provide a second phone-based session, this time focusing on facilitating personal choice (i.e., highlighting various treatment options the patient can choose from in addition to IOP). Otherwise, provide no further contact.''~\cite{Mckay2015} Notice that the DTR recommends intervention strategies for both engaged and non-engaged participants at week 8. Alternative names for DTRs include adaptive treatment strategies~\cite{Wallace2014,Ogbagaber2016} and adaptive interventions,~\cite{Almirall2014,Nahum-Shani2012a} among others.	
	
	Scientists often have questions about how best to sequence and individualize interventions in the context of a DTR. Sequential, multiple-assignment, randomized trials (SMARTs) are one type of randomized trial design that can be used to answer questions at multiple stages of the development of high-quality DTRs.\cite{Lavori2000,Lavori2004a,Murphy2005} The characteristic feature of a SMART is that some or all participants are randomized more than once, often based on previously-observed covariates. Each randomization corresponds to a critical question regarding the development of a high-quality DTR, typically related to the type, timing, or intensity of treatment. SMARTs have been employed in a variety of fields, including oncology,~\cite{Auyeung2009,Kidwell2014,Thall2016} surgery,~\cite{Diegidio2017,Hibbard2018} substance abuse,~\cite{Murphy2007} and autism.~\cite{Kasari2014}
	
	Most SMARTs contain ``embedded'' DTRs; that is, by design, participants in a SMART may be assigned to treatments which are consistent with recommendations made by one or more DTRs. The comparison of two embedded DTRs is a common primary aim for a SMART.~\cite{Nahum-Shani2012a} There exist data analytic methods for addressing this aim when the outcome is continuous,~\cite{Nahum-Shani2012a} survival,~\cite{Li2011} binary,~\cite{Kidwell2017} cluster-level~\cite{Necamp2017} and longitudinal.~\cite{Lu2016,Li2016} A key step in designing a SMART, as with any randomized trial, is determining the sample size needed to be able detect a desired effect with given power. However, there is no existing method for determining sample size for such a comparison when the outcome is continuous and longitudinal.
	
	The primary contribution of this manuscript is the development of tractable sample size formulae for SMARTs in which the primary aim is to compare two embedded DTRs using a continuous, longitudinal outcome. Additionally, we present estimators for parameters in the working covariance matrix used in the analysis methods developed by Lu et al.~\cite{Lu2016}
	
	In \cref{sec:SMARTs}, we provide a brief overview of three common SMART designs and introduce a motivating example. \Cref{sec:estimation} reviews the estimation procedure introduced by Lu et al., and extends it by developing estimators for various working covariance structures.\cite{Lu2016} In \cref{sec:sample-size}, we develop and present  sample size formulae for SMARTs in which the primary aim is a comparison of two embedded DTRs which recommend different first-stage treatments using a continuous repeated-measures outcome. The sample size formulae are evaluated via simulation in \cref{sec:sims}.

	\section{Dynamic Treatment Regimens and Sequential Multiple-Assignment Randomized Trials}
	\label{sec:SMARTs}
	A DTR is a sequence of functions (``decision rules''), each of which takes as inputs a person's history up to the time of the current decision (including baseline covariates, adherence, responses to previous treatments, etc.) and outputs a recommendation for the next treatment.~\cite{Murphy2005} Covariates which are used to recommend subsequent treatment are called ``tailoring variables''. Consider the example  DTR in \cref{sec:intro}. The recommended first-stage treatment is a phone-based session with a focus on re-engagement with the IOP. At week 8, each participant's history of engagement is assessed, and an appropriate second-stage treatment is recommended. For participants who have shown a pattern of continued non-engagement, the recommended second-stage treatment is a second phone-based session focusing on personal choice. For all other participants, the DTR recommends no further contact. The tailoring variable is an indicator as to whether or not the participant demonstrated a pattern of continued non-engagement prior to week 8. 
	
	We consider two-stage SMARTs in which the primary outcome is continuous and repeatedly measured in participants over the course of the study. Our examples refer to trials in which at least one observation of the outcome is made in each stage, though that is not required for the method presented here. For simplicity, we refer to the tailoring variable as response status to first-stage treatment, and, in the second stage, we describe participants as ``responders'' or ``non-responders''. We denote a DTR embedded in a SMART with a triple of the form $(a_{1}, a_{2R}, a_{2NR})$, where $a_{1}$ is an indicator for the recommended first-stage treatment, $a_{2R}$ an indicator for the second-stage treatment recommended for responders, and $a_{2NR}$ the second-stage treatment recommended for non-responders. 
	
	We introduce three common two-stage SMART designs in \cref{fig:all-smart-designs} which vary in the subsets of participants who are re-randomized after the first stage.

	\begin{figure}
		\centering
		\begin{subfigure}[t]{.48\textwidth}
			\resizebox{\textwidth}{!}{\input{smart-design--rerand-all.tikz}}
			\subcaption{All participants are re-randomized, regardless of response status.}
			\label[design]{fig:design-allrerand}
		\end{subfigure}
		\begin{subfigure}[t]{.48\textwidth}
			\resizebox{\textwidth}{!}{\input{smart-design--prototypical.tikz}}
			\subcaption{The second randomization is restricted to only non-responders.}
			\label[design]{fig:design-prototypical}
		\end{subfigure}
		\begin{subfigure}{.48\textwidth}
			\resizebox{\textwidth}{!}{\input{smart-design--autism.tikz}}
			\subcaption{The second randomization is restricted to only non-responders to treatment A.}
			\label[design]{fig:design-autism}
		\end{subfigure}
		\caption{Three commonly-used two-stage SMART designs. Each design varies in choice of which subsets of participants are re-randomized.}
		\label{fig:all-smart-designs}
	\end{figure}

	In \cref{sub@fig:design-allrerand}, all participants are re-randomized. There are eight DTRs embedded in this design: for example, the DTR which starts by recommending A, then recommends C for responders and F for non-responders. Using the notation in \cref{fig:all-smart-designs}, this DTR would be written (1, 1, -1). 
	SMARTs of this form have been run in the fields of drug dependence,~\cite{Oslin2005,Fitzsimons2015} smoking cessation,~\cite{Fu2017} and childhood depression,~\cite{Eckshtain2013} among others.
	
	SMARTs using \cref{sub@fig:design-prototypical} restrict the second randomization to only non-responders; that is, only participants who have a certain value of the tailoring variable (here, ``non-response'') are re-randomized. 
	This is perhaps the most common SMART design, and it has been utilized in the study of ADHD,~\cite{Pelham2016} adolescent marijuana use,~\cite{Budney2014} alcohol and cocaine dependence~\cite{Mckay2015}, and more. 
	There are four embedded DTRs in this design. Because responders are not re-randomized, $a_{2R}$ is set to zero for all embedded DTRs. 
	
	In \cref{sub@fig:design-autism}, re-randomization is restricted to only non-responders who receive a particular first-stage treatment. 
	SMARTs of this type have been used to investigate cognition in children with autism spectrum disorder~\cite{Kasari2014,Almirall2016} and implementation of a re-engagement program for patients with mental illness.~\cite{Kilbourne2013} 
	There are three DTRs embedded in this design. Note that, as in design II, responders are not re-randomized, so $a_{2R}$ is set to zero for all embedded DTRs. Furthermore, $a_{2NR}$ is set to zero when $a_{1} = -1$, as non-responders to treatment B are not re-randomized.
    
    	For more information on various SMART designs and case studies for each type, see Lei, et al.~\cite{Lei2012}
	
	To illustrate our ideas, we use ENGAGE, a SMART designed to study the effects of offering cocaine- and/or alcohol-dependent patients who did not engage in an IOP phone-based sessions either geared toward re-engaging them in an IOP or offering a choice of treatment options.~\cite{Mckay2015} The study recruited 500 cocaine- and/or alcohol-dependent adults who were enrolled in an IOP and failed to attend two or more sessions in the first two weeks. ENGAGE is modeled on \cref{sub@fig:design-prototypical}. In the context of \cref{fig:all-smart-designs}, treatment A was two phone-based motivational interviews focused on reengaging the participant with the IOP (``MI-IOP''); treatment B was two phone-based motivational interviews geared towards helping the participant choose and engage with an intervention of their choice (``MI-PC''). Participants who exhibited a pattern of continued non-engagement after eight weeks were considered non-responders, and re-randomized to receive either MI-PC (treatments D and G) or no further contact (treatments E and H). Responders were provided no further contact (treatments C and F). Following the coding in \cref{fig:all-smart-designs}, the example DTR from \cref{sec:intro} is labeled (1, 0, 1). 
	
	An important continuous outcome in ENGAGE is ``treatment readiness''. This is a measure of a patient's willingness and ability to commit to active participation in a substance abuse treatment program. The score ranges from 8-40 and is coded so that higher scores indicate greater treatment readiness. Measurements are taken at baseline, and 4, 8, 12, and 24 weeks after program entry.
		
	\section{Estimation}
	\label{sec:estimation}		
	We extend the work of Lu and colleagues by offering more detailed guidance on the estimation of model parameters used in computing quantities of interest on which to compare two embedded DTRs.~\cite{Lu2016} We first review the method below.
	
	\subsection{Marginal Mean Model}
	\label{sec:marginal-mean-model}	
	Consider a SMART design with embedded DTRs labeled by $(a_{1}, a_{2R}, a_{2NR})$. 
	Suppose we have a repeated-measures outcome $\bm{Y} = (Y_{t_{1}}, \ldots, Y_{t_{T}})$ observed such that $Y_{t}$ is measured for all participants at each of $T$ time points $\left\{t_{j} : j=1,\ldots, T; t_{1} < \ldots < t_{T}\right\}$. We do not require that these time points be equally-spaced. Define $t^*\in\left\{t_{j}\right\}$ to be the time of the measurement taken immediately before the second randomization. In ENGAGE, for example, $T=5$, $\left\{t_j\right\}= \{0, 4, 8, 12, 24\}$, and $t^*=t_{3}=8$. Let $\bm{X}$ be a vector of mean-centered baseline covariates, such as age at baseline, sex, etc.
	
	We are interested in $\E[Y_{t}^{(a_{1}, a_{2R}, a_{2NR})} \mid \bm{X}]$, the marginal mean of $\bm{Y}^{(a_{1}, a_{2R}, a_{2NR})}$ at time $t$ under DTR $(a_{1}, a_{2R}, a_{2NR})$ conditional on $\bm{X}$. This is the mean outcome at time $t$ had all individuals with characteristics $\bm{X}$ been offered DTR $(a_{1}, a_{2R}, a_{2NR})$. Recall that a DTR recommends treatments for both responders and non-responders; therefore, $\E[Y_{t}^{(a_{1}, a_{2R}, a_{2NR})} \mid \bm{X}]$ is marginal over response status.
	
	We impose a modeling assumption on $\E[Y_{t}^{(a_{1}, a_{2R}, a_{2NR})} \mid \bm{X}]$; namely, that $\E[Y_{t}^{(a_{1}, a_{2R}, a_{2NR})} \mid \bm{X}] = \mu_{t}^{(a_{1}, a_{2R}, a_{2NR})}\left(\bm{X}; \bm{\theta}\right)$, where $\mu_{t}^{(a_{1},a_{2R},a_{2NR})}\left(\bm{X}; \bm{\theta}\right)$ is a marginal structural mean model with unknown parameters $\bm{\theta} = (\bm{\eta}, \bm{\gamma})$. As discussed by Lu and colleagues, the sequential nature of treatment delivery in SMARTs may suggest constraints on the form of $\mu^{(a_{1}, a_{2R}, a_{2NR})}_{t}(\bm{X}; \bm{\theta})$.~\cite{Lu2016} The form of $\mu_{t}$ will depend, in part, on the design of the SMART. For instance, in ENGAGE, at time $t=0$, no treatments have been assigned, so all DTRs share a common mean. At times $t=4$ and $t=8$, the four embedded DTRs differ only by recommended first-stage treatment; thus there are two means of $Y_{t}^{(a_{1}, a_{2R}, a_{2NR})}$ at each timepoint. Finally, for times $t>t^*=8$, each DTR has a different mean $Y_{t}^{(a_{1}, a_{2R}, a_{2NR})}$. 
	
	An example marginal structural mean model for ENGAGE is
	\begin{equation}
	\label[model]{eq:designII-model}
		\begin{split}
			\mu_{t}^{(a_{1}, a_{2R}, a_{2NR})}(\bm{X};\bm{\theta}) ={}& \bm{\eta}^{\top}\bm{X} + \gamma_{0} + \mathbbm{1}_{\left\{t \leq t^*\right\}} \left(\gamma_{1} t + \gamma_{2}a_{1}t\right)  \\
			{}&+ \mathbbm{1}_{\left\{t>t^*\right\}} \left(t^*\gamma_{1} + t^*\gamma_{2}a_{1} + \gamma_{3}(t-t^*) + \gamma_{4}(t-t^*)a_{1} \right. \\
			{}& \phantom{+ \mathbbm{1}_{\left\{t>t^*\right\}} (} \left. + \gamma_{5} (t-t^*)a_{2NR} + \gamma_{6}(t-t^*)a_{1}a_{2NR}\right),
		\end{split}
	\end{equation}
	where $\mathbbm{1}_{\left\{E\right\}}$ is the indicator function for the event $E$.

	 Using contrast coding, i.e., $A_{i}\in\left\{-1,1\right\}$ for $i=1,2$, based on \cref{eq:designII-model}, 
	\begin{equation}
		\label{eq:interpretation-gamma2}
		2\gamma_{2} = E\left[\frac{Y_{t_{j}}^{(1, 0, \cdot)} - Y_{t_{k}}^{(1, 0, \cdot)}}{t_{j} - t_{k}} - \frac{Y_{t_{j}}^{(-1, 0, \cdot)} - Y_{t_{k}}^{(-1, 0, \cdot)}}{t_{j} - t_{k}} \mid \bm{X}\right], \quad t_{j},t_{k} \leq t^*,
	\end{equation}
	for example, represents the difference in slopes of expected treatment readiness in the first stage of the SMART between DTRs starting with different first-stage treatments. We present example models for \cref{sub@fig:design-allrerand,sub@fig:design-autism} in the online supplement. For more on modeling considerations for repeated-measures outcomes in SMARTs, see Lu et al.\cite{Lu2016}

	\subsection{Observed Data}
	\label{sec:obs-data}
	Suppose we have data arising from a SMART with $n$ participants. Let $A_{1,i} \in \left\{-1, 1\right\}$ denote the first-stage treatment randomly assigned to participant $i$, and let $R_{i} \in \left\{0,1\right\}$ indicate whether the $i$th participant responded to $A_{1,i}$, in which case $R_{i} = 1$, or not, so $R_{i} = 0$. Define $A_{2,i} \in \left\{-1, 1\right\}$ to be the randomly-assigned second-stage treatment. In \cref{sub@fig:design-prototypical}, since only non-responders are re-randomized, we set $A_{2, i} = 0$ for responders; similarly for \cref{sub@fig:design-autism}. We observe a continuous outcome $Y_{t, i}$ for each participant at each of $T$ timepoints. In general, the data collected on the $i$th individual over the course of the study are of the form 
	\[
		\left(\bm{X}_{i}, A_{1,i}, R_{i}, A_{2,i}, \bm{Y}_{i}\right),
	\]
	where $\bm{Y}_{i}$ is a length-$T$ vector consisting of all values of the outcome observed for the $i^{\text{th}}$ participant.
	
	\subsection{Estimating Equations}
	\label{sec:esteqns}
	Our goal is to estimate and make inferences on $\bm{\theta}$. Let $\mathcal{D}$ be the set of DTRs embedded in the SMART under study; for instance, in \cref{sub@fig:design-prototypical}, $\mathcal{D} = \left\{(a_{1}, a_{2R}, a_{2NR}) : a_{1} \in \left\{-1, 1\right\}, a_{2R} = 0, a_{2NR} \in \left\{-1, 1\right\} \right\}$.
	
	Let $W^{(d)}(A_{1,i}, R_{i}, A_{2,i})$ be a weight associated with participant $i$ and DTR $d\in\mathcal{D}$ defined as
	\begin{equation}
		\label{eq:weight-defn}
		W^{(d)}(A_{1,i}, R_{i}, A_{2,i}) =
		\frac{I^{(d)}(A_{1, i}, R_{i}, A_{2, i})}{P(A_{1,i} = a_{1})P(A_{2,i} = a_{2} \mid A_{1,i} = a_{1}, R_{i})},
	\end{equation}
	where $I^{(d)}(A_{1, i}, R_{i}, A_{2, i})$ is an indicator of whether participant $i$ is consistent with DTR $d$. Note that the form of $I^{(d)}(A_{1, i}, R_{i}, A_{2, i})$ depends on the particular SMART design under study; for each of the designs in \cref{fig:all-smart-designs}, these expressions are shown in \cref{tab:weights}. 
	$W^{(d)}(A_{1,i}, R_{i}, A_{2,i})$ is an inverse-probability-of-treatment weight used to correct for known imbalance in the proportion of responders and non-responders consistent with each DTR.~\cite{Nahum-Shani2012a,Cole2008,Chakraborty2013}
	In \cref{sub@fig:design-prototypical}, for example, only non-responders to first-stage treatment are re-randomized; if all randomizations are with probability 0.5, $W^{(1, 0, 1)}(1, 1, 0) = (.5\times 1)^{-1} = 2$ and $W^{(1, 0, 1)}(1, 0, 1) =(.5 \times .5)^{-1}= 4$. 
	
	\begin{table}
		\caption{Design-specific indicators for consistency with a given DTR $d \in \mathcal{D}$.}
		\label{tab:weights}
		\begin{tabular}{cc}
			\toprule
			Design & Form of $I^{(d)}$ \\ \midrule
			I   & $\mathbbm{1}_{\left\{A_{1, i} = a_{1}\right\}} (\mathbbm{1}_{\left\{A_{2, i} = a_{2R}\right\}}R_{i} + \mathbbm{1}_{\left\{A_{2, i} = a_{2NR}\right\}}(1-R_{i}))$ \\
			II  & $\mathbbm{1}_{\left\{A_{1, i} = a_{1}\right\}} (R_{i} + \mathbbm{1}_{\left\{A_{2, i} = a_{2NR}\right\}}(1-R_{i}))$ \\
			III & $\mathbbm{1}_{\left\{A_{1,i} = a_{1}\right\}} \left(\mathbbm{1}_{\left\{a_{1} = -1\right\}} + \mathbbm{1}_{\left\{a_{1} = 1\right\}}(R_{i} + \mathbbm{1}_{\left\{A_{2,i} = a_{2NR}\right\}}(1-R_{i}))\right)$ \\
			\bottomrule
		\end{tabular}
	\end{table}
	
	Define $\bm{D}^{(d)}\left(\bm{X}_{i}\right)$ to be the partial derivative of $\mu^{(d)}(\bm{X}_{i}; \bm{\theta})$ with respect to $\bm{\theta}^\top$. Let $\bm{V}^{(d)}(\bm{X}_{i}; \bm{\tau})$ be a working covariance matrix for $\bm{Y}^{(d)}$, conditional on baseline covariates $\bm{X}$, under DTR $d\in\mathcal{D}$; we discuss this quantity in detail in \cref{sec:workcov}. 
	We estimate $\bm{\theta}$ by solving the estimating equations 
	\begin{equation}
		\label{eq:esteqn}
		\begin{split}
			\bm{0} = \frac{1}{n} \sum_{i=1}^{n} \sum_{d \in \mathcal{D}} \left[W^{(d)}\left(A_{1,i}, R_{i}, A_{2,i}\right) \cdot \bm{D}^{(d)}(\bm{X}_{i})^\top \bm{V}^{(d)}(\bm{X}_{i}; \bm{\tau})^{-1} \left(\bm{Y}_{i} - \bm{\mu}^{(d)}(\bm{X}_{i}; \bm{\theta}\right) \right].
		\end{split}
	\end{equation}
	We call the solution to \cref{eq:esteqn} $\hat{\bm{\theta}}$.
	
	Under usual regularity conditions for $M$-estimators and given data from a SMART, $\hat{\bm{\theta}}$ is asymptotically consistent for $\bm{\theta}$ (see~\cref{sec:appendix:causal}). Furthermore, $\sqrt{n}\left(\hat{\bm{\theta}} - \bm{\theta}\right)$ has an asymptotic multivariate normal distribution:
	\[
		\sqrt{n}\left(\hat{\bm{\theta}} - \bm{\theta}\right) \Rightarrow \mathcal{N}\left(\bm{0}, \bm{B}^{-1} \bm{M} \bm{B}^{-1}\right),
	\]
	where 
	\begin{equation}
		\label{eq:bread-defn}
		\bm{B} := \E\left[\sum_{d\in\mathcal{D}} W^{(d)}\left(A_{1,i}, R_{i}, A_{2,i}\right) \bm{D}^{(d)}(\bm{X}_{i})^\top \bm{V}^{(d)}(\bm{X}_{i}; \bm{\tau})^{-1} \bm{D}^{(d)}(\bm{X}_{i}) \right]
	\end{equation}
	and
	\begin{equation}
		\label{eq:meat-defn}
		\bm{M} := \E\left[ \left(\sum_{d\in\mathcal{D}}  W^{(d)}\left(A_{1,i}, R_{i}, A_{2,i}\right) \bm{D}^{(d)}(\bm{X}_{i})^\top \bm{V}^{(d)}(\bm{X}_{i}; \bm{\tau})^{-1} \left(\bm{Y}_{i} - \bm{\mu}^{(d)}(\bm{X}_{i}; \bm{\theta})\right) \right)^{\otimes 2} \right],
	\end{equation}
	with $\bm{Z}^{\otimes 2} = \bm{ZZ}^\top$. Note that $\hat{\bm{\theta}}$ is consistent for $\bm{\theta}$ regardless of the choice of $\bm{V}^{(a_{1},a_{2})}(\bm{X}_{i}; \bm{\tau})$; however, we conjecture that choices closer to the true matrix will yield more efficient estimates.

	\subsection{Estimation of the Working Covariance Matrix}
	\label{sec:workcov}
	Decisions regarding the form of $\bm{V}^{(d)}(\bm{X}_{i}; \bm{\tau})$, $\bm{\tau} = (\bm{\sigma}, \bm{\rho})$, should be made by the scientist according to existing knowledge regarding the within-person correlation structure of $\bm{Y}^{(d)}$. In general, for an embedded DTR $d \in \mathcal{D}$, $\bm{V}^{(d)}(\bm{X}_{i}; \bm{\tau})$ takes the form
	\[
		\bm{V}^{(d)}(\bm{X}_{i}; \bm{\sigma}, {\bm{\rho}}) = \bm{S}^{(d)}(\bm{\sigma})^{1/2} \bm{R}^{(d)}(\bm{\rho}) \bm{S}^{(d)}(\bm{\sigma})^{1/2},
	\]
	where $\bm{S}^{(d)}(\bm{\sigma})^{1/2}\in \mathbb{R}^{T\times T}$ is a diagonal matrix with diagonal entries $\sigma^{(d)}_{t_1}, \ldots, \sigma^{(d)}_{t_T}$, and $\bm{R}^{(d)}(\bm{\rho})$ is a working correlation matrix for $\bm{Y}^{(d)}$. Note that this notation allows for different working covariance structures for each DTR, as well as non-constant variances in the repeated-measures outcome.
		
	We propose the following procedure to estimate $\bm{V}^{(d)}(\bm{X}_{i}; \bm{\tau})$. First, estimate $\bm{\theta}$ by solving~\cref{eq:esteqn} using the $T\times T$ identity matrix as $\bm{V}^{(d)}(\bm{X}_{i}; \bm{\tau})$ for all $d\in\mathcal{D}$. Call the solution $\hat{\bm{\theta}}_{(0)}$. Next, use $\hat{\bm{\theta}}_{(0)}$ to estimate $\sigma_{t}^{(d)}$ as follows
	\begin{equation}
		\label{eq:sigma-hat-not-marginalized}
		\hat{\sigma}^{(d)}_{t} = \frac{\sum_{i=1}^{n}{ W^{(d)}\left(A_{1, i}, R_{i}, A_{2,i}\right) \left(Y_{i,t} - \mu^{(d)}_{t}\left(\bm{X}_{i}; \hat{\bm{\theta}}^{(0)}\right)\right)^{2}}}{\sum_{i=1}^{n} W^{(d)}\left(A_{1, i}, R_{i}, A_{2,i}\right) - p},
	\end{equation}
	where $p$ is the dimension of $\bm{\theta}$. If the scientist believes that this variance is constant over time for each DTR, the estimate in \cref{eq:sigma-hat-not-marginalized} can be averaged over time; one can also average over DTR if one believes the variance is constant across all embedded DTRs. Estimators for $\bm{\rho}^{(d)}$ vary with choice of correlation structure $\bm{R}^{(d)}(\bm{\rho})$; we present estimators for selected structures in \cref{tab:cor-estimate-structures}. 
	Finally, to complete the estimation procedure, we again solve~\cref{eq:esteqn}, this time using $\hat{\bm{V}}^{(d)}(\bm{X}_{i}; \hat{\bm{\sigma}}, \hat{\bm{\rho}})$ as the working covariance matrix. This process can be further iterated, as suggested by Liang and Zeger; we call the final estimate of the model parameters $\hat{\bm{\theta}}$.~\cite{Liang1986}
	
	\begin{table}
		\caption{Correlation estimators for selected working correlation structures. The top entries define estimators under the assumption of constant within-person variance over time; the bottom entries allow for time-varying variances. $d\in\mathcal{D}$ is an embedded DTR, $W_{i}^{(d)}$ is shorthand for $W^{(d)}(A_{1,i}, R_{i}, A_{2, i})$, and $\hat{e}_{i,t}^{(d)}(\hat{\bm{\theta}})$ is the estimated residual $Y_{i, t} - \mu_{t}^{(d)}(\bm{X}_{i}; \hat{\bm{\theta}})$.}
		\label{tab:cor-estimate-structures}
		\begin{tabular}{ccc}
			\toprule
			Cor. structure & $\operatorname{Cor}(Y^{(d)}_{t_j}, Y^{(d)}_{t_k})$ & Estimator \\ \midrule
			AR(1) & $ \left\{
			\begin{array}{lr}
			1 & t_{j} = t_{k} \\
			\left(\rho^{(d)}\right)^{\left|j - k\right|} & t_{j} \neq t_{k}
			\end{array} \right.$ & 
			$\hat{\rho}^{(d)} = \frac{\sum_{i=1}^{n} W_{i}^{(d)} \sum_{m=1}^{T-1} \hat{e}_{i, t_m}^{(d)}(\hat{\bm{\theta}}) \hat{e}_{i, t_{m+1}}^{(d)}(\hat{\bm{\theta}})}{\left(\hat{\sigma}^{(d)}\right)^{2} \cdot n \cdot (T-1)}$ \\[15pt]
			Exchangeable & $ \left\{
			\begin{array}{lr}
				1 & t_j = t_k \\
				\rho^{(d)} & t_j \neq t_k
			\end{array} \right.$& 
			$\hat{\rho}^{(d)} = \frac{\sum_{i=1}^{n} W^{(d)}_{i} \sum_{l < m} \hat{e}_{i, t_l}^{(d)}(\hat{\bm{\theta}}) \hat{e}_{i, t_m}^{(d)}(\hat{\bm{\theta}})}{\left(\hat{\sigma}^{(d)}\right)^{2} \cdot n \cdot T(T-1)/2}$
			\\[15pt]
			Unstructured & $ \left\{
			\begin{array}{lr}
			1 & t_{j} = t_{k} \\
			\rho_{t_j,t_k}^{(d)} & t_{j} \neq t_{k}
			\end{array} \right.$ & 
			$\hat{\rho}_{t_j, t_k}^{(d)} = \frac{\sum_{i=1}^{n} W_{i}^{(d)}\hat{e}_{i, t_j}^{(d)}(\hat{\bm{\theta}}) \hat{e}_{i, t_k}^{(d)}(\hat{\bm{\theta}})}{\left(\hat{\sigma}^{(d)}\right)^{2} \cdot n}$\\  
			\bottomrule
		\end{tabular}
	\end{table}

	\section{Sample Size Formulae for End-of-Study Comparisons}
	\label{sec:sample-size}
	The estimation procedure presented in \cref{sec:estimation} is general. The marginal structural mean model $\mu^{(d)}(\bm{X}_{i}; \bm{\theta})$ can take any form appropriate for the SMART under analysis, data can be observed at any number of timepoints, and the working covariance matrix can have arbitrary structure.~\cite{Lu2016} 
    
    We now  sample size formulae for SMARTs in which the primary aim is to compare the mean end-of-study outcomes for two embedded DTRs that recommend different first-stage treatments and which satisfy certain design constraints. In particular, we restrict our focus to two-stage SMARTs in which the outcome is observed at three timepoints -- baseline, just prior to the second randomization, and at the end of the study -- and in which all randomizations occur with probability 0.5. Additionally, we consider a saturated, piecewise-linear mean structure $\mu^{(d)}(\bm{X}_{i}; \bm{\theta})$ similar to \cref{eq:designII-model}.
	
	Let $\bm{c}$ be some contrast vector so that the null hypothesis of interest takes the form
	\[
		H_{0}: \quad \bm{c}^\top \bm{\theta} = \bm{0},
	\]
	which we will test against an alternative of the form $H_{1}$: $\bm{c}^\top \bm{\theta} = \Delta$. To compare mean end-of-study outcomes between two embedded DTRs which recommend different first-stage treatments, the estimand of interest is
	\begin{equation}
		\label{eq:estimand}
		\bm{c}^\top\bm{\theta} = \E\left[Y_{2}^{(1,a_{\mathrm{2R}},a_{\mathrm{2NR}})} - Y_{2}^{(-1, a_{\mathrm{2R}}^\prime, a_{\mathrm{2NR}}^\prime)}\right],
	\end{equation}	
	for some choice of $a_{\mathrm{2R}}$, $a_{\mathrm{2R}}^\prime$, $a_{\mathrm{2NR}}$, and $a_{\mathrm{2NR}}^\prime$.
	For example, to test equality of mean end-of-study outcomes for DTRs (1, 0, 1) and (-1, 0, -1) in \cref{sub@fig:design-prototypical} under \cref{eq:designII-model} (assuming no $\bm{X}$, $\left\{t_{j}\right\} = \left\{0,1,2\right\}$, $t^*=1$), $\bm{c}$ is given by the linear combination $\bm{c}^\top \bm{\gamma}$, where $\bm{c}^\top = (0, 0, 2, 0, 2, 2, 0)$.
	
	We employ a 1-degree of freedom Wald test. The test statistic is
	\begin{equation*}
		Z = \frac{\sqrt{n} \bm{c}^\top \hat{\bm{\theta}}}{\sigma_{c}},
	\end{equation*}
	where $\sigma_{c} = \sqrt{\bm{c}^\top \bm{B}^{-1}\bm{M}\bm{B}^{-1}\bm{c}}$. Under the null hypothesis, by asymptotic normality of $\sqrt{n}\left(\hat{\bm{\theta}} - \bm{\theta}\right)$, the test statistic follows a standard normal distribution. 
	
	Define $\delta$ to be the standardized effect size as described by Cohen for an end-of-study comparison, i.e., 
	\begin{equation}
	\label{eq:effect-size-defn}
	\delta = \frac{\Delta}{\sigma},
	\end{equation}
	where $\sigma = \var(Y_{t}^{(d)})$ (see \cref{assume:exch-corr} below).~\cite{Cohen1988}
	
	The sample size formulae will require the response rate $P(R^{(a_{1})}=1) = r_{a_{1}}$. In order to simplify the form of $\sigma_c$ and obtain  tractable sample size formulae, we make two working assumptions:
	\begin{assumes}
		\item \textit{Constrained conditional covariance matrices for DTRs under comparison.} \label{assume:variances}
		\begin{assumes}
			\item \label{assume:leq-var} The variance in the outcome among non-responders after the second randomization is not too much larger than the corresponding variances in responders; in particular, for $t>t^*$ and $d\in\mathcal{D}$ under comparison, 
			\begin{multline*}
				\var\left(Y^{(d)}_{t} \mid R^{(a_{1})} = 0\right) - \var\left(Y^{(d)}_{t} \mid R^{(a_{1})} = 1\right) \\
				\leq (2-r_{a_{1}})\left(\E\left[Y_{t}^{(d)} \mid R^{(a_{1})} = 1\right] - \E\left[Y_{t}^{(d)} \mid R^{(a_{1})} = 0\right]\right)^2,
			\end{multline*}
			where $a_{1}$ is the first-stage treatment recommended by $d$ and $r_{a_{1}}$ is the probability of response to $a_{1}$. 
			
			\item \label{assume:eq-cov} The covariance between the end-of-study measurement and the measurements prior to the second stage among responders is less than or equal to the same quantity among non-responders:
			\[
				\operatorname{Cov}(Y^{(d)}_{t}, Y^{(d)}_{2} \mid R^{(a_{1})} = 1) \leq \operatorname{Cov}(Y^{(d)}_{t}, Y^{(d)}_{2} \mid R^{(a_1)} = 0)
			\]
			for all embedded DTRs $d\in\mathcal{D}$ and $t=0,1$. An additional, related assumption is given in \cref{sec:appendix:sample-size-derivation}.
		\end{assumes}
		\item \textit{Exchangeable marginal covariance structure.} \label{assume:exch-corr} The marginal variance of $\bm{Y}^{(d)}$ is constant across time and DTR, and has an exchangeable correlation structure with correlation $\rho$, i.e.,
		\[
			\var\left(\bm{Y}^{(d)}\right) = \bm{\Sigma} = \sigma^{2} \begin{bmatrix}
				1 & \rho & \rho \\
				\rho & 1 & \rho \\
				\rho & \rho & 1
			\end{bmatrix}
		\]
		for all $d\in\mathcal{D}$. 
	\end{assumes}
	
	Under \cref{assume:variances,assume:exch-corr}, the minimum-required sample size to detect a standardized effect size $\delta$ with power at least $1-\beta$ and two-sided type-I error $\alpha$ is
	\begin{equation}
		\label[formula]{eq:sample-size-EOS}
		n \geq \frac{4\left(z_{1-\alpha/2} + z_{1-\beta}\right)^{2}}{\delta^{2}} \cdot (1-\rho^{2}) \cdot \operatorname{DE},
	\end{equation}
	where $\operatorname{DE}$ is a SMART-specific ``design effect'' for an end-of-study comparison (see \cref{tab:design-effects}). 
	Note that the first term in \cref{eq:sample-size-EOS} is the typical sample size formula for a traditional two-arm randomized trial with a continuous end-of-study outcome and equal randomization probability. The middle term is due to the within-person correlation in the outcome, and is identical to the corresponding correction term for GEE analyses sized to detect a group-by-time interaction when there is no baseline group effect (see, e.g., Fitzmaurice et al., ch. 20\cite{Fitzmaurice2011}).
	
	The sample size formula presented in \cref{eq:sample-size-EOS} is conservative, particularly in settings in which $\rho$ is close to $(1+\sqrt{5})/2 \approx 0.62$. A sharper formula is available in \cref{sec:appendix:sample-size-derivation}; however, we emphasize \cref{eq:sample-size-EOS} as it is more immediately intepretable. We examine the performance of the sharp estimator in the supplement.
	
	\begin{table}
		\caption{Design effects for the sample size formulae in \cref{eq:sample-size-EOS}. $r_{a_{1}}$ is the response rate to first-stage treatment $a_{1}$}.
		\label{tab:design-effects}
		\begin{tabular}{ccc}
			\toprule
			 Design & Design effect & Conservative design effect \\
			 \midrule
			 I   & 2 & 2 \\
			 II  & $\frac{1}{2}(2-r_{1}) + \frac{1}{2}(2-r_{-1})$ & 2 \\
			 III & $\frac{1}{2}(3-r_{1})$ & $\frac{3}{2}$ \\
			 \bottomrule
		\end{tabular}
	\end{table}

	\Cref{assume:variances,assume:exch-corr} may be seen as overly simplifying; however, we will see in \cref{sec:sims,sec:discussion} that \cref{eq:sample-size-EOS} is robust to moderate violations of \cref{assume:variances} and that inputs to the formula can be adjusted in a way to accommodate violations of \cref{assume:exch-corr}. A working assumption similar to \labelcref{assume:leq-var} is commonly made in developing sample-size formulae for SMARTs using end-of-study outcomes.~\cite{Oetting2011,Kidwell2017,Necamp2017}
	\Cref{assume:eq-cov,assume:exch-corr} are necessary for the extension to the setting of a repeated-measures outcome. 
    
	\Cref{assume:variances} arises specifically as a consequence of unequal weights in \cref{eq:esteqn} (i.e., when there exists imbalance between responders and non-responders, by design); therefore, the assumption is not necessary in \cref{sub@fig:design-allrerand}, and can be relaxed to apply to only the two DTRs in which non-responders are re-randomized in \cref{sub@fig:design-autism}. See \cref{sec:appendix:sample-size-derivation} for more details on how this assumption is used. 
	Furthermore, \cref{assume:exch-corr} cannot be satisfied in \cref{sub@fig:design-allrerand} if all eight embedded DTRs have unique means. We investigate robustness to this in \cref{sec:sims}.
	
	\section{Simulations}
	\label{sec:sims}
	
	We performed a variety of simulations to assess the performance of the proposed estimators and sample size formula. In particular, we focus on four types of scenarios: first, when no assumptions are violated; second, when each of \cref{assume:leq-var,assume:eq-cov,assume:exch-corr} are violated. In each scenario, we are interested in the empirical power of a comparison of the DTR which recommends only treatments indicated by $1$ and the DTR which recommends only treatments indicated by $-1$. Sample sizes are computed based on nominal power $1-\beta = 0.8$ and two-sided type-1 error $\alpha = 0.05$. Each empirical result is based on 5000 simulated data sets.
	
	For each simulation, the true marginal mean model is as in \cref{eq:designII-model} for \cref{sub@fig:design-prototypical}; analogous models are used for \cref{sub@fig:design-allrerand,sub@fig:design-autism} -- see the supplement for examples. We do not include baseline covariates $\bm{X}$; this is a conservative approach, as adjustment for prognostic covariates typically will increase power.~\cite{Kahan2014} Estimates of marginal means from ENGAGE were used to inform a reasonable range of ``true'' means from which to simulate, though the scenarios presented here are not designed to mimic ENGAGE exactly. All simulations take $T=3$ and values of $\bm{\gamma}$ are chosen to achieve $\delta = 0.3$ or $\delta = 0.5$ (``small'' and ``moderate'' effect sizes, respectively). Data were generated according to a conditional mean model which, when averaged over response, yields the marginal model of interest. Outcomes $\bm{Y}_{i}$ were simulated from a multivariate normal distribution with means suggested by the conditional model, and covariance matrices which, when averaged over response, produce the DTR-specific marginal variance structure of the form in \cref{assume:exch-corr}. Additional details of the generative model used for simulations can be found in \cref{sec:appendix:sims}. 
	
	Simulated data sets were analyzed using the method described in \cref{sec:estimation}, using an exchangeable working covariance structure with correlation $\rho$ and variance $\sigma^2$ in all scenarios. $\rho$ and $\sigma$ are treated as common across time and DTR for the purposes of estimation. The estimation procedure for $\bm{\gamma}$ and $\bm{\tau}$ described in \cref{sec:estimation} was iterated until the norms of the estimates were within $10^{-8}$ of the previous estimates' norms. 

	\subsection{Simulation Results}

	\begin{table}
		\caption{Sample sizes and empirical power results for an end-of-study comparison of the DTR recommending only treatments indexed by $1$ and that which recommends only treatments indicated by $-1$. $\delta$ is the standardized effect size defined in \cref{eq:effect-size-defn}. Bolded results are significantly less than 0.8 at the 0.05 level.}
		\label{tab:sims1}
        \begin{small}
		\begin{tabular}{ccccccccc}
			\toprule & & & & & \multicolumn{4}{c}{Empirical power} \\ \cmidrule{6-9}
			& & & & & & \multicolumn{2}{c}{Violation of \labelcref{assume:variances}} & {Violation of \labelcref{assume:exch-corr}} \\
			\cmidrule{7-9}
			Design & $\delta$ & $r$ & $\rho$ & $n$ &  \labelcref{assume:variances,assume:exch-corr} satisfied & \labelcref{assume:leq-var} & \labelcref{assume:eq-cov} & True AR(1) \\ \midrule 
			\labelcref{sub@fig:design-allrerand} 
			  & 0.3 & 0.4 & 0   & 698 & 0.797 & 0.800 & --    & -- \\
			  &     &     & 0.3 & 635 & 0.807 & 0.811 & 0.820 & \textbf{0.778} \\
			  &     &     & 0.6 & 447 & 0.842 & 0.829 & 0.830 & \textbf{0.712} \\
			  &     &     & 0.8 & 252 & 0.848 & 0.838 & 0.844 & \textbf{0.662} \\
			  &     & 0.6 & 0   & 698 & 0.816 & 0.794 & --    & --  \\
			  &     &     & 0.3 & 635 & 0.825 & 0.801 & 0.813 & \textbf{0.778} \\
			  &     &     & 0.6 & 447 & 0.829 & 0.833 & 0.833 & \textbf{0.723} \\
			  &     &     & 0.8 & 252 & 0.851 & 0.832 & 0.838 & \textbf{0.665} \\
			  & 0.5 & 0.4 & 0   & 252 & 0.804 & 0.810 & --    & -- \\
			  &     &     & 0.3 & 229 & 0.818 & 0.812 & 0.820 & \textbf{0.783} \\
			  &     &     & 0.6 & 161 & 0.843 & 0.829 & 0.837 & \textbf{0.710} \\
			  &     &     & 0.8 & 91  & 0.845$^*$ & 0.840$^*$ & 0.840$^*$ & \textbf{0.676}$^*$ \\
			  &     & 0.6 & 0   & 252 & 0.809 & 0.797 & --    & -- \\
			  &     &     & 0.3 & 229 & 0.816 & 0.818 & 0.812 & \textbf{0.777} \\
			  &     &     & 0.6 & 161 & 0.838 & 0.831 & 0.831 & \textbf{0.713} \\
			  &     &     & 0.8 & 91  & 0.853 & 0.840 & 0.846 & \textbf{0.666} \\
			\midrule
			\labelcref{sub@fig:design-prototypical}
			  & 0.3 & 0.4 & 0   & 559 & 0.800 & \textbf{0.790} & --             & --  \\
			  &     &     & 0.3 & 508 & 0.803 & \textbf{0.786} & \textbf{0.785} & \textbf{0.757} \\
			  &     &     & 0.6 & 358 & 0.824 & 0.795          & \textbf{0.779} & \textbf{0.695} \\
			  &     &     & 0.8 & 201 & 0.825 & \textbf{0.785} & 0.803          & \textbf{0.625} \\
			  &     & 0.6 & 0   & 489 & 0.796 & \textbf{0.773} & --             & -- \\
			  &     &     & 0.3 & 445 & 0.797 & \textbf{0.787} & \textbf{0.786} & \textbf{0.767} \\
			  &     &     & 0.6 & 313 & 0.812 & \textbf{0.783} & \textbf{0.766} & \textbf{0.679} \\
			  &     &     & 0.8 & 176 & 0.827 & \textbf{0.756} & \textbf{0.774} & \textbf{0.625} \\
			  & 0.5 & 0.4 & 0   & 201 & 0.794 & 0.793          & --             & -- \\
			  &     &     & 0.3 & 183 & 0.815 & 0.794          & \textbf{0.789} & \textbf{0.774} \\
			  &     &     & 0.6 & 129 & 0.830 & 0.797          & 0.793          & \textbf{0.699} \\
			  &     &     & 0.8 & 73  & 0.839 & \textbf{0.787} & 0.807          & \textbf{0.638} \\
			  &     & 0.6 & 0   & 176 & 0.806 & \textbf{0.765} & --             & -- \\
			  &     &     & 0.3 & 160 & 0.815 & \textbf{0.773} & 0.802          & \textbf{0.778} \\
			  &     &     & 0.6 & 113 & 0.816$^*$ & \textbf{0.773} & \textbf{0.763} & \textbf{0.691} \\
			  &     &     & 0.8 & 64  & 0.831$^*$ & \textbf{0.775} & \textbf{0.787} & \textbf{0.643} \\
			  \midrule
			  \labelcref{sub@fig:design-autism}
			  & 0.3 & 0.4 & 0   & 454 & 0.798 & 0.793 & -- & -- \\
			  &     &     & 0.3 & 413 & 0.805 & 0.800 & 0.800 & \textbf{0.760} \\
			  &     &     & 0.6 & 291 & 0.808 & 0.803 & 0.797 & \textbf{0.677} \\
			  &     &     & 0.8 & 164 & 0.825 & 0.800 & 0.802 & \textbf{0.611} \\
			  &     & 0.6 & 0   & 419 & 0.798 & 0.805 & -- & --  \\
			  &     &     & 0.3 & 381 & 0.802 & 0.793 & 0.795 & \textbf{0.753} \\
			  &     &     & 0.6 & 268 & 0.814 & 0.803 & \textbf{0.786} & \textbf{0.686} \\
			  &     &     & 0.8 & 151 & 0.824 & 0.794 & \textbf{0.784} & \textbf{0.611} \\
			  & 0.5 & 0.4 & 0   & 164 & 0.802 & \textbf{0.790} & -- & -- \\
			  &     &     & 0.3 & 149 & 0.814 & 0.803 & 0.805 & \textbf{0.773} \\
			  &     &     & 0.6 & 105 & 0.815 & 0.807 & 0.796 & \textbf{0.683} \\
			  &     &     & 0.8 &  59 & 0.811 & 0.815$^*$ & 0.817$^*$ & \textbf{0.635}$^*$ \\
			  &     & 0.6 & 0   & 151 & 0.792 & 0.791 & -- & -- \\
			  &     &     & 0.3 & 138 & 0.813 & 0.802 & 0.799 & \textbf{0.769} \\
			  &     &     & 0.6 &  97 & 0.818$^*$ & 0.804$^*$ & 0.796$^*$ &  \textbf{0.690}$^*$ \\
			  &     &     & 0.8 &  55 & 0.824$^*$ & 0.797$^*$ & 0.797$^*$ & \textbf{0.630}$^*$ \\
			\bottomrule
            
		\end{tabular}
        \end{small} \\
		\footnotesize{\raggedright{$^*$ Fewer than 5000 simulations generated data in which all treatment sequences were observed.}}
	\end{table}
	
	Simulation results are compiled in \cref{tab:sims1}. We find that the sample size formula presented in \cref{eq:sample-size-EOS} performs as expected when all assumptions are satisified. Empirical power is not significantly less than the target power of 0.8, per a one-sided binomial test with level $0.05$. The sample size is, as expected, slightly conservative, particularly when within-person correlation is high. There may be some concern that, for high within-person correlation, \cref{eq:sample-size-EOS} is overly conservative; should this concern arise, we recommend use of the sharper formulae presented in the supplement.
	
	Violation of \cref{assume:leq-var} was induced by lowering the end-of-study variance among responders relative to that among non-responders, while keeping the marginal variance fixed. In particular, the results shown in \cref{tab:sims1} correspond to approximately a 25\% reduction in responders' variance relative to the non-responders' variance minus the correction in \cref{assume:leq-var} for all DTRs. 
	
	As conjectured in \cref{sec:sample-size}, violating \cref{assume:leq-var} does not impact empirical power in \cref{sub@fig:design-allrerand}, since the assumption arises as a consequence of imbalanced numbers of responders and non-responders consistent with a particular DTR (see \cref{sec:appendix:sample-size-derivation}). For \cref{sub@fig:design-prototypical}, empirical power is consistently less than the nominal value when \cref{assume:leq-var} is violated. However, while the empirical power is often significantly less than 0.8, the observed loss of power is relatively small. For \cref{sub@fig:design-autism}, we notice small reductions in power relative to scenarios in which both \cref{assume:variances,assume:exch-corr} are satisfied, though the conservative nature of \cref{eq:sample-size-EOS} appears to protect against more severe loss of power. This suggests that our sample size formula is moderately robust to ``reasonable'' violations of \labelcref{assume:leq-var}. 
	
	Violation of \cref{assume:eq-cov} was induced by choosing $\operatorname{Cor}(Y^{(d)}_{t}, Y^{(d)}_{2} \mid R^{(a_{1})} = 1) > \operatorname{Cor}(Y^{(d)}_{t}, Y^{(d)}_{2} \mid R = 0)$ while keeping respective variances fixed. There exist natural constraints on how much larger than $\operatorname{Cov}(Y^{(d)}_{t}, Y^{(d)}_{2} \mid R = 0)$ the responders' covariance can be while ensuring that (1) all conditional covariance matrices are positive definite and (2) $\operatorname{Cov}(Y^{(d)}_{t}, Y^{(d)}_{2} \mid R = 0) \geq 0$ for $t=0,1$. These constraints vary with $\rho$. The empirical power results shown in \cref{tab:sims1} were generated by choosing $\operatorname{Cor}(Y^{(d)}_{t}, Y^{(d)}_{2} \mid R^{(a_{1})} = 1)$ such that $\operatorname{Cov}(Y^{(d)}_{t}, Y^{(d)}_{2} \mid R^{(a_{1})} = 1)$ is the midpoint between the minimum covariance for which the assumption is violated and the maximum covariance allowed by the aforementioned constraints. Simulation results show that our sample size formula is quite robust to violations for low-to-moderate within-person correlations; at high correlations, the empirical power is significantly less than 0.8. However, as with \cref{assume:leq-var}, the observed reduction in power is not unreasonable. Furthermore, when within-person correlation is high, sample size becomes rather small. Since the method presented here is based on asymptotic normality, we caution the reader that small sample sizes (e.g., $n<100$) provided by \cref{eq:sample-size-EOS} may be quite sensitive to violation of the working assumptions.
	
	The final columns of \cref{tab:sims1} suggest that \cref{eq:sample-size-EOS} is highly sensitive to violations of \cref{assume:exch-corr} in regards to the true correlation structure. In particular, when the true correlation structure is not exchangeable with correlation $\rho$ and is instead AR(1) with correlation $\rho$, empirical power is substantially lower than the target of 0.8, particularly as $\rho$ increases. This is unsurprising: under an AR(1) correlation structure, less information about the end-of-study outcome is provided by, say, the baseline measure than under an exchangeable correlation structure. Since, by using \cref{eq:sample-size-EOS}, we have assumed more information is available from earlier measurements than is actually the case, we will be underpowered. As our assumed $\rho$ increases, the difference between the assumed and actual correlation between the end-of-study measurement and earlier measurements increases, leading to more severe loss of power.
	
	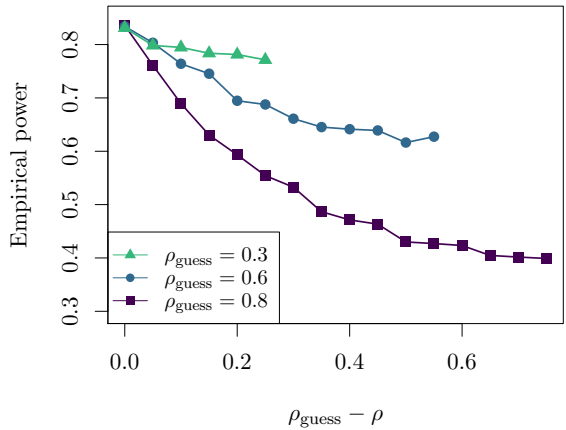
\begin{figure}
		\centering
		\resizebox{.5\textwidth}{!}{\input{violS2RhoR4Delta3.tikz}}
		\caption{Empirical power versus the difference between the true within-person correlation $\rho$ and hypothesized correlation $\rho_{\mathrm{guess}}$ used to compute sample size. Results are shown for \cref{sub@fig:design-prototypical} with a hypothesized response rate of 0.4, and sample size was chosen to detect standardized effect size $\delta = 0.3$ for the comparison of DTRs (1, 0, 1) and (-1, 0, -1). Each point is based on 5000 simulations with target power $0.8$ and significance level $0.05$. Results are extremely similar for \cref{sub@fig:design-allrerand,sub@fig:design-autism} and different values of $\delta$ and $r$ (see supplement).}
		\label{fig:misspecified-rho}
	\end{figure}
	
	In \cref{fig:misspecified-rho}, we examine the effect on empirical power of misspecifying the within-person correlation. Analytically, we see from \cref{eq:sample-size-EOS} that if the assumed $\rho$ is smaller than the true within-person correlation, the sample size will be conservative. On the other hand, when the assumed $\rho$ in \cref{eq:sample-size-EOS} is larger than the true correlation, the sample size will be anti-conservative. \Cref{fig:misspecified-rho} shows plots of empirical power against the difference between the assumed within-person correlation $\rho_{\mathrm{guess}}$ and the true $\rho$. For small $\rho_{\mathrm{guess}}$, \cref{eq:sample-size-EOS} appears to be quite robust to misspecification of $\rho$; however, as $\rho_{\mathrm{guess}}$ increases, the formula becomes highly sensitive to such a violation of \cref{assume:exch-corr}. This is supported analytically, since \cref{eq:sample-size-EOS} is a function of $\rho_{\mathrm{guess}}^2$.
	
	\section{Discussion}
    \label{sec:discussion}
	We have derived sample size formulae for SMART designs in which the primary aim is a comparsion of two embedded DTRs that begin with different first-stage treatments on a continuous, repeated-measures outcome. We derived the formulae for three common SMART designs.
	
	The sample size formula is the product of three components: (1) the formula for the minimum sample size for the comparison of two means in a standard two-arm trial (see, e.g., Friedman et al.,~\cite{Friedman2010} page 147), (2) a deflation factor of $1-\rho^2$ that accounts for the use of a repeated-measures outcome, and (3) a SMART-specific ``design effect'', an inflation factor that accounts for the SMART design.

	The SMART design effect can be interpreted as the cost of conducting the SMART relative to conducting a standard two-arm randomized trial of the two DTRs which comprise the primary aim. The benefit of conducting a SMART (relative to the standard two-arm randomized trial) is the ability to answer additional, secondary questions that are useful for constructing effective DTRs. For example, such questions may focus on one or more of the other pairwise comparisons between DTRs, on whether the first- and second-stage treatments work synergistically to impact outcomes (e.g., a test of the null that $\gamma_6=0$ in \cref{eq:designII-model}), or may focus on hypothesis-generating analyses that seek to estimate more deeply-tailored DTRs.~\cite{Watkins1989,Nahum-Shani2012,Zhang2015}

	The formulae are expected to be easy-to-use for both applied statistical workers and clinicians. Indeed, inputs $\alpha$, $\beta$, and $\Delta$ are as in the sample size formula for a standard $z$-test. Furthermore, estimates of $\rho$, $r_{a_1}$, and $\sigma$ are often readily available from the literature or can be estimated using data from prior studies (e.g., prior randomized trials, or external pilot studies). 

	We make a number of recommendations concerning the use of the formulae; in particular, how best to use the formulae conservatively in the absence of certainty concerning prior estimates of $\rho$, $r_{a_1}$, and/or the structure of the variance of the repeated measures outcome. First, in \cref{sub@fig:design-prototypical,sub@fig:design-autism}, if there is uncertainty concerning the response rate (e.g., response rate estimates are based on data from smaller prior studies), one approach is to err conservatively by assuming a smaller-than-estimated response rate. In both designs, the most conservative approach is to assume a response rate of zero.
	
	Second, as in standard randomized trials in which the primary aim is a pre-post comparison, the required sample size decreases as the hypothesized within-person correlation increases.~\cite{Zhang2014} Therefore, if the hypothesized $\rho$ is larger than the true $\rho$, the computed sample size will be anti-conservative, resulting in an under-powered study. Indeed, we see this in the results of the simulation experiment (see \cref{fig:misspecified-rho}). Here, again, one approach is to err conservatively towards smaller values of $\rho$.  
	
	Finally, \cref{assume:exch-corr} (concerning the variance of the repeated measures outcome) may be seen as overly restrictive in the imposition of an exchangeable correlation structure. For example, studies with a continuous repeated measures outcome may observe an autoregressive correlation structure.  However, the exchangeable working assumption can be employed conservatively in the following way: If the hypothesized structure is not exchangeable, one approach is to set $\rho$ in \cref{eq:sample-size-EOS} to the smallest plausible value (e.g., the within-person correlation between the baseline and end-of-study measurements for an autoregressive structure). Because this approach utilizes a lower bound on the value of  the true within-person correlations, it is expected to yield a larger than needed (more conservative) sample size. 
	Similarly, if the true within-person correlation is expected to differ by DTR, one approach is to employ the smallest plausible $\rho$. As with the third recommendation, these recommendations are not unique to SMARTs; indeed, these strategies may also be used to size standard two-arm randomized trials with repeated measures outcome.
	
	In the case where $\var(Y_{t}^{(d)})$ varies with time and/or DTR, we conjecture that power will suffer if a pooled estimate of $\sigma^2$ is used when the variance decreases with time. To see this, consider that the standardized effect size $\delta$ defined in \cref{eq:effect-size-defn} has as a denominator the pooled standard deviation of $Y_{2}^{(d)}$ across the groups under comparison. Should the estimate of pooled standard deviation  be larger than the true value, the variance of $\bm{c}^\top\hat{\bm{\theta}}$ will increase; since the estimate will be less efficient than hypothesized, power will be lower than expected. Conversely, we also conjecture that when $\var(Y_t^{(d)})$ increases with $t$, the sample size will be conservative using similar reasoning. 

	There are a number of interesting ways to build on this manucript in future methodological work.  First, some scientists may be interested in a primary aim comparison that involves other features of the marginal mean trajectory, such as the area under the curve (AUC). Future work could develop formulae for these other primary aim comparisons.  An important challenge here is in whether and how to define the standardized effect size $\delta$.
	Second, an interesting extension of this work is to better understand the cost-benefit trade-off between adding additional sample size versus adding additional measurement occasions to the SMART design. The formulae presented here employ the rather simplistic working assumption that there are $T=3$ measurement occasions (at baseline, the end of the first stage, and the end of the second stage). Based on limited simulation experiments, sample sizes based on our formulae are expected to perform conservatively when $T>3$. Future work could develop rules of thumb for how best to allocate additional sample size versus additional measurement occasions given budget constraints (e.g., a fixed total study cost and fixed costs for an additional participant and additional measurement occasion).  
	Third, as the field moves toward simulation-based approaches for sample size calculation, there is a clear need for the development of software that would allow applied statistical workers and clinicians to make fewer (or more flexible) assumptions concerning many of the features of the SMART, or to be more flexible with respect to the design of the SMART.  An important challenge here is to make the software general enough to be used across a number of different types of SMART designs (e.g., three stages of randomization), yet not so flexible that it is difficult to use. The benefits of this is the ability to examine the power for various different scientific questions given a single data generative model and for many other types of SMARTs.

	\begin{acks}
		This work was supported by the Eunice Kennedy Shriver National Institute of Child Health and Human Development [grant
		number R01HD073975]; the National Institute of Biomedical Imaging and Bioengineering [grant number U54EB020404]; the
		National Institute of Mental Health [grant number R03MH097954, R01MH114203]; the National Institute on Alcohol Abuse
		and Alcoholism [grant numbers P01AA016821, RC1AA019092]; the National Institute on Drug Abuse [grant numbers
		R01DA039901, P50DA039838, R01DA047279]; the National Cancer Institute [grant number U01CA229437]; and the
		Institute of Education Sciences [grant number R324B180003]. This research was supported in part through computational
		resources and services provided by Advanced Research Computing at the University of Michigan, Ann Arbor.
	\end{acks}

	\bibliographystyle{unsrtnat}
	\bibliography{rmSMART-Sample-Size}
	
	\newpage
	
	\appendix
	\renewcommand{\theequation}{A\arabic{equation}}
	\setcounter{equation}{0}
	
	\section{Identifiability Assumptions}
	\label{sec:appendix:causal}
	We make the following assumptions in order to show that \cref{eq:esteqn} has mean zero. 
	\begin{assumec}
		\item \textit{Positivity.} \label{assume:positivity} The probabilities $P(A_1=1)$ and $P(A_2=1 \mid A_1, R )$ are non-zero. 		
        \item \textit{Consistency.\cite{Robins1997}} \label{assume:consistency} A participant's observed responder status is consistent with the participant's corresponding potential responder status under the assigned first-stage treatment; i.e., $R_{i} = \ind{A_{1,i} = 1} R^{(1)} + \ind{A_{1,i} = -1} R^{(-1)}$. And a participant's observed  repeated measures outcomes are consistent with the participant's corresponding potential repeated measures outcomes under the assigned treatment sequence; see Table~\ref{tab:consistency}. 
		
		\begin{table}
			\caption{Design-specific consistency assumptions. $d\in\mathcal{D}$ indexes embedded DTRs $(a_{1}, a_{2R}, a_{2NR})$.}
			\label{tab:consistency}
			\begin{tabular}{cc}
				\toprule
				Design & $\bm{Y}_{i}$ equals \\ \midrule
				I & $\sum_{d\in\mathcal{D}} \frac{1}{2}\ind{A_{1,i} = a_{1}} \left(R_{i}\ind{A_{2,i} = a_{2R}} + (1-R_{i}) \ind{A_{2,i} = a_{2NR}}\right) \bm{Y}^{(d)}$ \\
				II & $\sum_{d\in\mathcal{D}} \ind{A_{1,i}=a_{1}} \left(\frac{1}{2}R_{i} + (1-R_{i})\ind{A_{2,i}=a_{2}}\right) \bm{Y}^{(d)}$ \\
				III & $\sum_{d\in\mathcal{D}} \ind{A_{1,i} = a_{1}} \left(\ind{a_{1} = -1} + \ind{a_{1} = 1}\left(\frac{1}{2}R_{i} + (1-R_{i})\ind{A_{2,i}=a_{2}}\right)\right)\bm{Y}^{(d)}$ \\
				\bottomrule
			\end{tabular}
			\\ The factor of $1/2$ for responders in designs II and III accounts for the fact that these participants are consistent with two DTRs. For example in design II, if $R_{j} = 1$ for some $j$, $Y_{j}^{(a_{1}, 1)} = Y_{j}^{(a_{1}, -1)} := Y_{j}^{(a_{1}, 0)}$.
		\end{table}
		
		\item \textit{Sequential randomization.} \label{assume:seq-rand} At each stage in the SMART, observed treatments $A_1$ and $A_2$ are assigned independently of future potential outcomes, given the participant's history up to that point. That is,
		\begin{gather*}
	    \{\bm{Y}^{(d)}, R(a_1) \} \perp\!\!\!\perp A_{1}    \\
		\{\bm{Y}^{(d)} \} \perp\!\!\!\perp A_{2} \mid A_{1},  R   
%		A_{1} \perp\!\!\!\perp \bm{Y}^{(a_{1}, a_{2R}, a_{2NR})}, R.
		\end{gather*}
	\end{assumec}

	\Cref{assume:positivity,assume:seq-rand} are satisfied by design in a SMART (see, e.g., Lavori and Dawson\cite{Lavori2014}); \cref{assume:consistency} is connects the potential outcomes and observed data, and is typically accepted in the analysis of randomized trials.

	\section{Description of Simulation Framework}
	\label{sec:appendix:sims}
	
	For each simulation experiment, data were generated as follows:
	\begin{enumerate}
		\item Draw $A_{1}$ from $\left\{-1, 1\right\}$ with equal probability.
		\item For each subset of participants, generate $R$ from a Bernoulli($r_{A_{1}}$) distribution.
		\item Draw $A_{2}$ from $\left\{-1, 1\right\}$ with equal probability for those participants who are re-randomized (e.g., for all participants in \cref{sub@fig:design-allrerand} but only for non-responders in \cref{sub@fig:design-prototypical}).
		\item For each participant, generate $\bm{Y}\in \mathbb{R}^{T\times 1}$ such that $\bm{Y} = \bm{\nu}^{(A_{1}, R, A_{2})}(\bm{\gamma}, \bm{\lambda}) + \bm{\epsilon}^{(A_{1}, R, A_{2})}$, where $\bm{\epsilon}^{(A_{1}, R, A_{2})} \sim \bm{\mathcal{N}}_{T}(\bm{0}, \bm{\Sigma}^{(A_{1}, R, A_{2})})$; we define $\bm{\Sigma}^{(A_{1}, R, A_{2})})$ below.
	\end{enumerate}
	
	The estimands of primary interest are marginal over response to first-stage treatment; however, the simulation framework described above uses generative models which are conditional on response to first-stage treatment. Here, we describe how to derive conditional quantities which, when averaged over response, yield the desired marginal characteristics. 
	
	We first describe the conditional mean models used to generate $\bm{Y}$ in step 4. For all designs, the conditional mean model used is
	\begin{equation}
		\nu_{t}^{(A_{1}, R, A_{2R})}(\bm{\gamma}, \bm{\lambda}) = \mu_{t}^{\left(A_{1}, R \cdot A_{2R}, (1-R) \cdot A_{2NR}\right)}(\bm{\gamma}) + \mathbbm{1}_{\left\{t>t^*\right\}}(t-t^*) (R-r_{A_{1}}) (\lambda_{1} + \lambda_{2} A_{1}),
		\label{eq:cond-model}
	\end{equation}
	where $\mu_{t}^{\left(A_{1}, R \cdot A_{2R}, (1-R) \cdot A_{2NR}\right)}(\bm{\gamma})$ is the design-specific marginal mean model (see the supplement for examples). 
	Notice the absence of $\bm{X}$; since our sample size formulae do not account for baseline covariates, we omit them for our simulations. Denote by $\bm{\nu}^{(A_{1}, R, A_{2R})}(\bm{\gamma}, \bm{\lambda})$ the length-$T$ vector which has $j^{\text{th}}$ element $\nu_{t_{j}}^{(A_{1}, R, A_{2R})}(\bm{\gamma}, \bm{\lambda})$.

	In \cref{eq:cond-model}, we can interepret $\lambda_{1}$ as the average difference in ``second-stage intercept'' between responders and non-responders to the same first-stage treatment. Similarly, we can view $\lambda_{2}$ as the average difference in ``second-stage intercept'' between responders and non-responders to opposite first-stage treatments. Together, $\bm{\gamma}$ and $\bm{\lambda}$ completely specify all conditional means for each design which, by construction, are guaranteed to average over response to $\mu_{t}^{\left(a_{1}, a_{2R}, a_{2NR}\right)}(\bm{\gamma})$ for all $t > t^*$. 
	
	It remains to compute conditional variance matrices $\bm{\Sigma}^{(A_{1}, R, A_{2})}$. As with the mean models, these conditional covariance matrices must account for the design of the SMART, which may suggest constraints on the possible forms $\bm{\Sigma}^{(A_{1}, R, A_{2})}$ can take (see \cref{sec:marginal-mean-model} or Lu, et al.~\cite{Lu2016}). In particular, the ``conditional'' variances of $Y_{0}$ and $Y_{1}$ are exactly the marginal variances, since response has not yet been observed. Under \cref{assume:exch-corr}, we have
	\[
		\bm{\Sigma}^{(A_{1}, R, A_{2})} = 
		\begin{bmatrix}
			\sigma^{2} & \rho \sigma^{2} & \sigma \rho_{02}^{(A_1, R, A_2)} \sigma_{2}^{(A_1, R, A_2)} \\
			\rho \sigma^2 & \sigma^2 & \sigma \rho_{12}^{(A_1, R, A_2)} \sigma_{2}^{(A_1, R, A_2)} \\
			\sigma \rho_{02}^{(A_1, R, A_2)} \sigma_{2}^{(A_1, R, A_2)} & \sigma \rho_{12}^{(A_1, R, A_2)} \sigma_{2}^{(A_1, R, A_2)} & \left(\sigma_{2}^{(A_1, R, A_2)}\right)^2
		\end{bmatrix},
	\]
	
	We assume, for simplicity in our generative model, that $\rho_{02}^{(A_1, R, A_2)} = \rho_{12}^{(A_1, R, A_2)} = \rho^{(A_1, R, A_2)}$ for all $A_{1}$, $R$, and $A_{2}$. Given these quantities for responders, i.e., $\sigma_{2}^{(A_{1}, 1, A_{2})}$ and $\rho^{(A_1, 1, A_2)}$, we can find appropriate values for the non-responders such that the conditional variance matrices marginalize correctly. By the law of total variance, we have 
	\[
		\left(\sigma_{2}^{(A_1, 0, A_2)}\right)^{2} = \frac{1}{1-r_{A_{1}}}\cdot \left(\sigma^{2} - r_{A_{1}}\left(\sigma_{2}^{(A_1, 1, A_2)}\right)^2\right) - r_{A_{1}}\left(\nu_{2}^{A_{1}, 1, A_{2}}(\bm{\gamma}, \bm{\lambda}) - \nu_{2}^{A_{1}, 0, A_{2}}(\bm{\gamma}, \bm{\lambda})\right)^{2}.
	\]
	Similarly, by the law of total covariance, we have
	\[
		\rho^{(A_{1}, 0, A_{2})} = \frac{\rho\sigma - r_{A_{1}} \rho^{(A_{1}, 1, A_{2})}\sigma_{2}^{(A_{1}, 1, A_{2})} }{(1-r_{A_{1}})\sigma_{2}^{(A_1, 0, A_2)}}.
	\]
	This fully specifies the data generative model discussed above.
	
	\section{Derivation of Sample Size Formulae}
	\label{sec:appendix:sample-size-derivation}
	We derive the sample size formulae for comparing two DTRs which recommend different first-stage treatments that are embedded in a SMART in which a continuous repeated-measures outcome is collected throughout the study. These formulae are based on the regression analyses described in~\cref{sec:estimation} and a Wald test. 
	
	We consider a SMART in which the outcome is collected three timepoints: at baseline $(t=0)$, immediately before assessing response/non-response $(t=1)$, and at the end of the study $(t=2)$. We ignore the presence of baseline covariates $\bm{X}$ and assume $\bm{\mu}^{(d)}(\bm{\theta})$ is piecewise-linear in $\bm{\theta}$ (see, for example, \cref{eq:designII-model}). 
	
	Recall from \cref{sec:sample-size} that we wish to the null hypothesis $H_{0}: \bm{c}^\top\bm{\theta} = 0$. In particular, we are interested in contrasts $\bm{c}$ which yield an end-of-study comparison between two embedded DTRs which recommend different first-stage treatments.  Since a comparison of two embedded DTRs will yield a 1-degree of freedom Wald test, we use a $Z$ statistic:
	\begin{equation*}
		Z = \frac{\sqrt{n} \bm{c}^\top \hat{\bm{\theta}}}{\sigma_{c}},
	\end{equation*}
	where $\sigma_{c} = \sqrt{\bm{c}^\top \bm{B}^{-1}\bm{M}\bm{B}^{-1}\bm{c}}$. Under $H_{0}$, by asymptotic normality of $\sqrt{n}(\hat{\bm{\theta}} - \bm{\theta})$, the test statistic follows an asymptotic standard normal distribution. If we wish to size the SMART to detect the alternative hypothesis $\bm{c}^\top \bm{\theta} = \Delta$, we arrive at the following form for the minimum-required sample size:
	\begin{equation}
		\label[formula]{eq:sample-size-general2}
		n \geq \left(z_{1-\alpha/2} + z_{1-\beta}\right)^{2} \frac{\sigma^{2}_{c}}{\Delta^{2}},
	\end{equation}
	where $z_{p}$ is the $p$th quantile of the standard normal distribution. \Cref{eq:sample-size-general2} is a fairly standard result in the clinical trials literature;~\cite{Lachin1981,Friedman2010} however, because of the dependence on $\sigma_c$, the formula is not useful as written. The goal of this appendix is to derive a closed-form upper bound on $\sigma_c$ so as to obtain a sample size formula in terms of marginal quantities which can be more easily elicited from clinicians, or estimated from the literature.  
	
	Recall the definitions of $\bm{B}$ and $\bm{M}$ in \cref{eq:bread-defn,eq:meat-defn}, respectively. These quantities depend on $\bm{D}^{(d)}$, the partial derivative matrix of $\bm{\mu}^{(d)}(\bm{\theta})$ and $\bm{V}^{(d)}(\bm{\tau})$, the working covariance matrix for $\bm{Y}$. By assumed linearity of $\bm{\mu}^{(d)}(\bm{\theta})$, $\bm{D}^{(d)}$ is a fixed, constant matrix for all $d$. Furthermore, we assume that the working covariance matrix $\bm{V}^{(d)}(\bm{\tau})$ is correctly specified and satisifies \cref{assume:exch-corr} so that $\bm{V}^{(d)}(\bm{\tau}) = \bm{\Sigma}$ for all $d \in \mathcal{D}$.
	
	The estimand in \cref{eq:estimand} is a function of potential outcomes; as written in \cref{eq:bread-defn,eq:meat-defn}, $\bm{B}$ and $\bm{M}$ are functions of observed data. We begin by expressing $\bm{B}$ in terms of potential outcomes. Under the positivity, consistency, and sequential ignorability conditions (\cref{assume:positivity,assume:consistency,assume:seq-rand}), we can apply lemma 4.1 of Murphy et al.~\cite{Murphy2001} so that
	\begin{align*}
		\bm{B} &= \sum_{d\in\mathcal{D}} \E_{A_{1}, R, A_{2}}\left[ W^{(d)}(A_{1}, R, A_{2}) \bm{D}^{(d)} \left(\bm{V}^{(d)}(\bm{\tau})\right)^{-1}\left(\bm{D}^{(d)}\right)^\top \right] \\
		&= \sum_{d\in\mathcal{D}} \bm{D}^{(d)} \bm{\Sigma}^{-1}\left(\bm{D}^{(d)}\right)^\top. \numbereqn \label{eq:expanded-bread}
	\end{align*}
	
	We now turn our attention to $\bm{M}$. Expanding the outer product inside the expectation, we have
	\begin{multline}
		\bm{M} ={} \E_{A_{1}, R, A_{2}, \bm{Y}} \left[ \left(\sum_{d\in\mathcal{D}}W^{(d)}(A_{1}, R, A_{2}) \bm{D}^{(d)} \left(\bm{V}^{(d)}(\bm{\tau})\right)^{-1} \left(\bm{Y} - \bm{\mu}^{(d)}(\bm{\theta})\right) \right)^{\otimes 2} \right]  \\
		\begin{split}
			\label{eq:expanded-meat}
			={}& \sum_{d\in\mathcal{D}} \E_{A_{1}, R, A_{2}, \bm{Y}}\left[\left(W^{(d)}(A_{1}, R, A_{2}) \right)^2 \left(\bm{D}^{(d)} \left(\bm{V}^{(d)}(\bm{\tau})\right)^{-1} \left(\bm{Y} - \bm{\mu}^{(d)}(\bm{\theta})\right)\right)^{\otimes 2}\right] \\
			&{}+ \sum_{d\neq d^\prime} \E_{A_{1}, R, A_{2}, \bm{Y}}\left[W^{(d)}(A_{1}, R, A_{2}) W^{(d^\prime)}(A_{1}, R, A_{2}) \bm{D}^{(d)} \left(\bm{V}^{(d)}(\bm{\tau})\right)^{-1} \right.  \\
			& \phantom{=+\sum_{d\neq d^\prime} \E_{A_{1}, R, A_{2}, \bm{Y}}} {} \left. \left(\bm{Y} - \bm{\mu}^{(d)}(\bm{\theta})\right) \left(\bm{Y} - \bm{\mu}^{(d^\prime)}(\bm{\theta})\right)^\top \left(\bm{V}^{(d^\prime)}(\bm{\tau})\right)^{-1} \left(\bm{D}^{(d^\prime)}\right)^\top \right].
		\end{split}
	\end{multline}
	
	Notice that the work above is design-independent: $\bm{B}$ and $\bm{M}$ have the same form as \cref{eq:expanded-bread,eq:expanded-meat}, respectively, for all designs. Below, we proceed only for \cref{sub@fig:design-prototypical}, but derivations for \cref{sub@fig:design-allrerand,sub@fig:design-autism} are analogous, substituting appropriate definitions of $W^{(d)}(A_{1},R, A_{2})$. Recall that, for \cref{sub@fig:design-prototypical}, when all randomization probabilities are 0.5,  $W^{(d)}(A_{1}, R, A_{2}) = 2\mathbbm{1}_{\left\{ A_{1} = 1 \right\}} (R + 2(1-R)\mathbbm{1}_{\left\{ A_{2} = 1 \right\}})$.
	
	Consider a single summand of the first term in \cref{eq:expanded-meat}; for concreteness, choose (without loss of generality) $d = 1$, which we will say corresponds to the DTR which recommends only treatments indicated by 1. Again applying lemma 4.1 of Murphy et al.,~\cite{Murphy2001} we have	
	
	\begin{align}
		&\E_{A_{1}, R, A_{2}, \bm{Y}}\left[\left(W^{(1)}(A_{1}, R, A_{2}) \right)^2 \left(\bm{D}^{(1)} \left(\bm{V}^{(1)}(\bm{\tau})\right)^{-1} \left(\bm{Y} - \bm{\mu}^{(1)}(\bm{\theta})\right)\right)^{\otimes 2}\right] \nonumber \\
		&\phantom{=}= \E_{R^{(1)}, \bm{Y}^{(1)}} \left[2 \left(R^{(1)} + 2\left(1 - R^{(1)}\right)\right) \left(\bm{D}^{(1)} \bm{\Sigma}^{-1} \left(\bm{Y}^{(1)} - \bm{\mu}^{(1)}(\bm{\theta})\right)\right)^{\otimes 2}\right] \label{eq:meat-derivation-weight}\\
		\begin{split}
			\label{eq:meat-derivation-smoothing}
			&\phantom{=}={} 2r_{1}\E_{\bm{Y}^{(1)} \mid R^{(1)}=1} \left[\left(\bm{D}^{(1)} \bm{\Sigma}^{-1}\left(\bm{Y}^{(1)} - \bm{\mu}^{(1)}(\bm{\theta})\right)\right)^{\otimes 2} \mid R^{(1)} = 1\right]  \\
			&\phantom{==}+{} 4(1-r_{1})\E_{\bm{Y}^{(1)} \mid R^{(1)}=0} \left[\left(\bm{D}^{(1)} \bm{\Sigma}^{-1}\left(\bm{Y}^{(1)} - \bm{\mu}^{(1)}(\bm{\theta})\right)\right)^{\otimes 2} \mid R^{(1)} = 0\right] 
		\end{split} \\
		&\phantom{=}= \bm{D}^{(1)} \bm{\Sigma}^{-1} \left(2r_{1} \bm{\Sigma}^{(1, 1, 0)} + 4(1-r_1) \bm{\Sigma}^{(1, 0, 1)}\right) \bm{\Sigma}^{-1} \left(\bm{D}^{(1)}\right)^\top, \label{eq:meat-derivation-condvar-expansion}
	\end{align}
	where $\bm{\Sigma}^{(A_{1}, R, A_{2})}$ is defined as in \cref{sec:appendix:sims}. \Cref{eq:meat-derivation-weight} is a consequence of the definition of the weight and \cref{assume:consistency};
	\cref{eq:meat-derivation-smoothing} follows by smoothing.
	
	We now construct an upper bound on $\sigma_c$ using marginal quantities. Recall from \cref{sec:sample-size} that, for an end-of-study comparison in \cref{sub@fig:design-prototypical} with $\left\{t_{j}\right\} = \left\{0, 1, 2\right\}$ and $t^*=1$, the appropriate contrast vector for the test is $\bm{c}^\top = (0, 0, 2, 0, 2, 2, 0)$. Under \cref{assume:exch-corr}, for the DTRs under comparison, we find that
	\begin{equation}
		\begin{split}
			\label{eq:meat-cis-sandwich-comp}
			&\bm{c}^\top \bm{B}^{-1}\bm{D}^{(1)} \bm{\Sigma}^{-1} \left(2(2-r_{1})\bm{\Sigma} - 2r_{1} \bm{\Sigma}^{(1, 1, 0)} + 4(1-r_1) \bm{\Sigma}^{(1, 0, 1)}\right) \bm{\Sigma}^{-1} \left(\bm{D}^{(1)}\right)^\top\bm{B}^{-1}\bm{c} \\
			& \phantom{=}= \frac{2}{1+\rho} \left[(1+\rho)\left((2-r_1)\sigma^2 - 2(1-r_{1})\left(\sigma_{2}^{(1,0,1)}\right)^2 + r_{1}\left(\sigma_{2}^{(1,1,1)}\right)^2\right)\right. \\
			& \phantom{=} \phantom{= \frac{2}{1+\rho} + } + \rho(2+\rho)\left( -(2-r_{1})\rho\sigma^2 + 2(1-r_{1})\rho_{02}^{(1,0,1)}\sigma\sigma_{2}^{(1,0,1)} + r\rho_{02}^{(1,1,1)}\sigma\sigma_{2}^{(1,1,1)}\right) \\
			& \phantom{=} \phantom{= \frac{2}{1+\rho} + } + \left. \rho \left(-(2-r_1)\rho\sigma^2 + 2(1-r_1)\rho_{12}^{(1,0,1)}\sigma\sigma_{2}^{(1,0,1)} + r_{1}\rho_{12}^{(1,1,1)}\sigma\sigma_{2}^{(1,1,1)}\right)\right].
		\end{split}
	\end{equation}
	Under \cref{assume:variances}, \cref{eq:meat-cis-sandwich-comp} is non-negative. Indeed, notice that the first line of the right-hand side of \cref{eq:meat-cis-sandwich-comp} is positive under \cref{assume:leq-var}; the second and third lines are positive under \cref{assume:eq-cov}. 
	For DTRs not under comparison, we have
	\begin{multline}
		\label{eq:meat-cis-sandwich-other}
		\bm{c}^\top \bm{B}^{-1}\bm{D}^{(1)} \bm{\Sigma}^{-1} \left(2(2-r_{1})\bm{\Sigma} - 2r_{1} \bm{\Sigma}^{(1, 1, 0)} + 4(1-r_1) \bm{\Sigma}^{(1, 0, 1)}\right) \bm{\Sigma}^{-1} \left(\bm{D}^{(1)}\right)^\top\bm{B}^{-1}\bm{c} \\
		\phantom{=}= \frac{2}{1+\rho} \sigma^{2}(2-r_{a_{1}})\left(1+\rho-2\rho^2\right),
	\end{multline}
	which is non-negative for all $\rho\in[0,1]$. Thus, for all $d\in\mathcal{D}$, we have
	\begin{multline}
		\label{eq:cis-meat-bound}
		\bm{c}^\top\bm{B}^{-1}\bm{D}^{(d)} \bm{\Sigma}^{-1} \left(2r_{a_1} \bm{\Sigma}^{(a_1, 1, 0)} + 4(1-r_1) \bm{\Sigma}^{(a_1, 0, a_{2NR})}\right) \bm{\Sigma}^{-1} \left(\bm{D}^{(d)}\right)^\top\bm{B}^{-1}\bm{c}^\top \\
		\leq 2\left(2-r_{a_1}\right)\bm{c}^\top\bm{B}^{-1}\bm{D}^{(d)} \bm{\Sigma}^{-1} \left(\bm{D}^{(d)}\right)^\top\bm{B}^{-1}\bm{c}.
	\end{multline}
	This establishes an upper bound on the first term of \cref{eq:expanded-meat}.
	
	Now, in the second term of \cref{eq:expanded-meat}, notice that any product of the form $W^{(1, a_{2})}(A_1,R,A_2) \cdot W^{(-1, b_{2})}(A_1,R,A_2)$ is identically zero (recall \cref{eq:weight-defn} and \cref{tab:weights}), so we consider only products of DTRs which start with the same first-stage treatment. Again by \cref{eq:weight-defn}, $W^{(a_{1}, 1)}(A_1,R,A_2) \cdot W^{(a_{1}, -1)}(A_1,R,A_2) = 4\mathbbm{1}_{\left\{A_{1} = a_{1}\right\}}R$. Consider, as above, a single summand; in particular, the product for DTRs (1, 0, 1) and (1, 0, -1), indexed by $d=1$ and $d=2$, respectively. By \cref{assume:consistency,assume:positivity} and steps similar to those above, we 
	\begin{multline}
		\label{eq:meat-crossprod-expand}
		\E_{A_{1}, R, \bm{Y}}\left[4 \mathbbm{1}_{\left\{A_1 = 1\right\}} R \bm{D}^{(1)} \left(\bm{V}^{(1)}(\bm{\tau})\right)^{-1} \left(\bm{Y} - \bm{\mu}^{(1)}(\bm{\theta})\right) \left(\bm{Y} - \bm{\mu}^{(2)}(\bm{\theta})\right)^\top \left(\bm{V}^{(2)}(\bm{\tau})\right)^{-1} {\bm{D}^{(2)}}^\top \right] \\
		\phantom{=}= 2r_1\bm{D}^{(1)} \bm{\Sigma}^{-1} \E_{\bm{Y}^{(1)}, \bm{Y}^{(2)} \mid R^{(1)}} \left[\left(\bm{Y}^{(1)} - \bm{\mu}^{(1)}(\bm{\theta})\right) \left(\bm{Y}^{(2)} - \bm{\mu}^{(2)}(\bm{\theta})\right)^\top \mid R^{(1)} = 1\right] \bm{\Sigma}^{-1} \left(\bm{D}^{(2)}\right)^\top .
	\end{multline}
	
	Recall that, by design, $Y^{(d)}_{0}$ does not depend on $d$ and $Y_{1}^{(d)}$ does not depend on $a_{2}$. Thus, under \cref{assume:exch-corr}, we have
	\begin{equation}
	\label{eq:meat-crossprod-conclusion}
	\E_{\bm{Y}^{(1)}, \bm{Y}^{(2)} \mid R^{(1)}} \left[ \left(\bm{Y}^{(1)} - \bm{\mu}^{(1)}(\bm{\theta})\right) \left(\bm{Y}^{(2)} - \bm{\mu}^{(1)}(\bm{\theta})\right)^\top \mid R^{(1)} = 1\right] 
	=
	\bm{\Sigma}^{(1,1,0)} + \bm{C}_{1},
	\end{equation}
	where $\bm{C}_{a_1}$ is a $T\times T$ matrix with $(T,T)$ element $\left(\nu_{2}^{(a_1,1,0)}(\bm{\theta}) - \mu_{2}^{(a_1,0,1)}(\bm{\theta})\right)\left(\nu_{2}^{(a_1,1,0)}(\bm{\theta}) - \mu_{2}^{(a_1,0,-1)}(\bm{\theta})\right)$ and all other entries zero.
	
	As above, we seek to find an upper bound for \cref{eq:meat-crossprod-expand} involving marginal quantities. For any two DTRs $d,d^\prime$ that share responders,
	\begin{multline*}
		2r_{a_1}\bm{c}\bm{B}^{-1}\bm{D}^{(d)}\bm{\Sigma}^{-1}\left(\bm{\Sigma} - \bm{\Sigma}^{(a_{1},1,0)}+\bm{C}_{a_1}\right)\bm{\Sigma}^{-1}\left(\bm{D}^{(d^\prime)}\right)^\top \bm{B}^{-1}\bm{c}^\top \\
		= \frac{r_{a_1}}{1+\rho} \left(\rho\left(\rho^2\sigma^2 + \rho_{12}^{(a_{1},1,0)}\sigma\sigma_{2}^{(a_{1},1,0)} - \rho\sigma(\sigma - \rho_{02}^{(a_{1}, 1, 0)}\sigma_2^{(a_{1}, 1, 0)}\right)\right),
	\end{multline*} 
	which can be shown to be non-negative under \cref{assume:eq-cov}, provided the difference between $\operatorname{Cov}(Y^{(d)}_{1}, Y^{(d)}_{2} \mid R^{(a_{1})=1}$ and $\rho\operatorname{Cov}(Y^{(d)}_{0}, Y^{(d)}_{2} \mid R^{(a_{1})=1}$ is not too large. Therefore, for all $d,d^\prime \in \mathcal{D}$, 
	\begin{multline}
		\label{eq:trans-meat-bound}
		2r_{a_1}\bm{c}\bm{B}^{-1}\bm{D}^{(d)}\bm{\Sigma}^{-1}\left(\bm{\Sigma}^{(a_{1},1,0)}+\bm{C}_{a_1}\right)\bm{\Sigma}^{-1}\left(\bm{D}^{(d^\prime)}\right)^\top \bm{B}^{-1}\bm{c}^\top \\
		\leq 2r_{a_1}\bm{c}^\top\bm{B}^{-1}\bm{D}^{(d)} \bm{\Sigma}^{-1} \left(\bm{D}^{(d^\prime)}\right)^\top\bm{B}^{-1}\bm{c}.
	\end{multline}
	
	We now use the upper bounds computed in \cref{eq:trans-meat-bound,eq:cis-meat-bound} and arrive at the conclusion that for the saturated model in \cref{eq:designII-model}, if we wish to compare DTRs $(1,0,1)$ and $(-1,0,-1)$ using an end-of-study outcome, 
	\begin{equation}
		\sigma_{c}^{2} = \bm{c}^\top \bm{B}^{-1} \bm{M} \bm{B}^{-1} \bm{c} \leq \frac{4 \sigma^2 (1-\rho)  \left(\rho^2+4 \rho - \frac{1}{2}(r_{1}+r_{-1}) (2 \rho+1)+2\right)}{1+\rho}
		\label{eq:sample-size-sharp-d2}
	\end{equation}
	Plugging \cref{eq:sample-size-sharp-d2} into \cref{eq:sample-size-general2} leads to the aforementioned ``sharp'' sample size formula for \cref{sub@fig:design-prototypical}. Some algebra shows that 
	\begin{equation}
		\sigma_{c}^{2} \leq 4\sigma^{2}\cdot\left(1-\rho^2\right)\cdot \frac{1}{2}\left((2-r_1)+(2-r_{-1})\right),
	\end{equation}
	which allows for an easy-to-understand sample size formula. Plugging this result into \cref{eq:sample-size-general2}, we arrive at \cref{eq:sample-size-EOS}.
	
	\section{Code Repository}
	\label{sec:code-repo}
	The R Code used to generate the simulation results in this paper can be obtained from
    \begin{center}
    	\url{https://github.com/nseewald1/rmSMARTSize}. 
    \end{center}
\end{document}

%% file: smart-design--rerand-all.tikz
\begin{tikzpicture}[%
	node distance=6mm,
	randomize/.style={
		circle,
		minimum size=6mm,
		thick,
		black,
		draw
	},
	rerand/.style={
		xshift = 15mm
	},
	treatment/.style={
		rectangle,
		minimum size = 7mm,
		thick,
		black,
		anchor=west,
		align=center,
		draw
	},
	blank/.style={
		rectangle,
		minimum size=2mm
	},
	subgroup/.style={
		rectangle,
		rounded corners=1.5mm,
		thin,
		minimum size=6mm,
		black,
		draw
	},
	rlabel/.style={
		above,
		xshift=-4mm,
		align=left
	},
	nrlabel/.style={
		below,
		xshift=-1mm,
		align=left
	},
	trialarrow/.style={
		thick,
		decoration={markings,mark=at position 1 with  {\arrow[scale=1.5,>=stealth]{>}}},
		postaction={decorate}
	},
	subsetarrow/.style={
		thin,
		decoration={markings,mark=at position 1 with {\arrow[scale=1.2,>=stealth]{>}}},
		postaction={decorate}
	}, thick]
					
	\matrix[row sep=-2mm,column sep=12mm] { %
		& \node [align=center,xshift=6mm] {\large \textbf{Stage 1}}; & & & \node [align=center,xshift=5mm] {\large \textbf{Stage 2}}; \\[5mm]
		
		% DESIGN I
		
		& & & & \node (C-I) [treatment] {C \\ $A_{2R} = 1$};\\
		& & \node (b1-I) [blank] {}; & \node (R2-I) [randomize, rerand] {R}; & \\
		& & & & \node (D-I) [treatment] {D \\ $A_{2R} = -1$};\\
		& \node (A-I) [treatment] {A \\ $A_{1} = 1$}; & & & \\
		& & & & \node (E-I) [treatment] {E \\ $A_{2NR} = 1$}; \\
		& & \node (b2-I) [blank] {}; &s \node (R3-I) [randomize, rerand] {R}; & \\
		& & & & \node (F-I) [treatment] {F \\ $A_{2NR} = -1$}; \\
		\node (R1-I) [randomize] {R}; & & & &\\
		& & & & \node (G-I) [treatment] {G \\ $A_{2} = 1$}; \\
		& & \node (b3-I) [blank] {}; & \node (R4-I) [randomize, rerand] {R}; & \\
		& & & & \node (H-I) [treatment] {H \\ $A_{2} = -1$}; \\
		& \node (B-I) [treatment] {B \\ $A_{1} = -1$}; & & &\\
		& & & & \node (I-I) [treatment] {I \\ $A_{2} = 1$};\\
		& & \node (b4-I) [blank] {}; & \node (R5-I) [randomize, rerand] {R}; & \\
		& & & & \node (J-I) [treatment] {J \\ $A_{2} = -1$}; \\[1cm]
		
		% TIMEPOINTS
		
		\node (time0) [align=center, anchor=center] {Time 0};& & & \node (time1) [align=center, anchor=center, xshift=15mm] {Time 1}; & \node (time2) [align=center, xshift = 10mm] {Time 2}; \\
	};

%		\draw[dashed,lightgray] (time1.north) -- (R2-III.south);
%		\draw[dashed,lightgray] (R2-III.north) -- (R3-II.south);
%		\draw[dashed,lightgray] (R3-II.north) --++(90:3.2cm);
%		
%		\draw[dashed,lightgray] (time2.north west) --++(90:33cm);

	% DESIGN I LINES

	\draw[trialarrow] (R1-I) -- (A-I.west);
	\draw (A-I.east) -- (b1-I.center);
	\draw (A-I.east) -- (b2-I.center);
	\draw[trialarrow] (b1-I.center) -- node[rlabel] {Responders} (R2-I.west);
	\draw[trialarrow] (b2-I.center) -- node[nrlabel] {Non-Responders} (R3-I.west);
	\draw[trialarrow] (R2-I.east) -- (C-I.west);
	\draw[trialarrow] (R2-I.east) -- (D-I.west);
	\draw[trialarrow] (R3-I.east) -- (E-I.west);
	\draw[trialarrow] (R3-I.east) -- (F-I.west);
	
	\draw[trialarrow] (R1-I) -- (B-I.west);
	\draw (B-I.east) -- (b3-I.center);
	\draw (B-I.east) -- (b4-I.center);
	\draw[trialarrow] (b3-I.center) -- node[rlabel] {Responders} (R4-I.west);
	\draw[trialarrow] (b4-I.center) -- node[nrlabel] {Non-Responders} (R5-I.west);
	\draw[trialarrow] (R4-I.east) -- (G-I.west);
	\draw[trialarrow] (R4-I.east) -- (H-I.west);
	\draw[trialarrow] (R5-I.east) -- (I-I.west);
	\draw[trialarrow] (R5-I.east) -- (J-I.west);
	
	% TIME LINES
	
	\draw (time0) -- (time1) -- (time2);
\end{tikzpicture}

%% file: smart-design--prototypical.tikz
\begin{tikzpicture}[%
	node distance=6mm,
	randomize/.style={
		circle,
		minimum size=6mm,
		thick,
		black,
		draw
	},
	rerand/.style={
		xshift = 15mm
	},
	treatment/.style={
		rectangle,
		minimum size = 7mm,
		thick,
		black,
		anchor=west,
		align=center,
		draw
	},
	blank/.style={
		rectangle,
		minimum size=2mm
	},
	subgroup/.style={
		rectangle,
		rounded corners=1.5mm,
		thin,
		minimum size=6mm,
		black,
		draw
	},
	rlabel/.style={
		above,
		xshift=1mm,
		align=center
	},
	nrlabel/.style={
		below,
		xshift=-1mm,
		align=left
	},
	trialarrow/.style={
		thick,
		decoration={markings,mark=at position 1 with  {\arrow[scale=1.5,>=stealth]{>}}},
		postaction={decorate}
	},
	subsetarrow/.style={
		thin,
		decoration={markings,mark=at position 1 with {\arrow[scale=1.2,>=stealth]{>}}},
		postaction={decorate}
	}, thick]
					
	\matrix[row sep=-2mm,column sep=12mm] { %
		& \node [align=center,xshift=6mm] (stage1) {\large \textbf{Stage 1}}; & & & \node [align=center,xshift=5mm] (stage2) {\large \textbf{Stage 2}};\\[5mm]
		%
		%
		% DESIGN II
		
		\node (designII) [blank] {}; 
		& & \node (b1-II) [blank] {}; & \node (b2-II) [blank] {}; & \node (C-II) [treatment] {C \\ $A_{2} = 0$};\\
		& \node (stage1align) [treatment,white] {}; & & & \\
		& & & & \node (D-II) [treatment] {D \\ $A_{2} = 1$};\\
		& & \node (b3-II) [blank] {}; & \node (R2-II) [randomize, rerand] {R}; & \\
		& & & & \node (E-II) [treatment] {E \\ $A_{2} = -1$};\\			\node [randomize,white] {};  & & & & \\
		& & \node (b4-II) [blank] {}; & \node (b5-II) [blank] {}; & \node (F-II) [treatment] {F \\ $A_{2} = 0$};\\
		& \node [treatment, white] {}; & & & \\
		& & & & \node (G-II) [treatment] {G \\ $A_{2} = 1$};\\
		& & \node (b6-II) [blank] {}; & \node (R3-II) [randomize, rerand] {R}; & \\
		& & & & \node (H-II) [treatment] {H \\ $A_{2} = -1$};\\[1cm]
		
		% TIMEPOINTS
		
		\node (time0) [align=center, anchor=center] {Time 0};& & & \node (time1) [align=center, anchor=center, xshift=15mm] {Time 1}; & \node (time2) [align=center, xshift = 10mm] {Time 2}; \\
	};

%		\draw[dashed,lightgray] (time1.north) -- (R2-III.south);
%		\draw[dashed,lightgray] (R2-III.north) -- (R3-II.south);
%		\draw[dashed,lightgray] (R3-II.north) --++(90:3.2cm);
%		
%		\draw[dashed,lightgray] (time2.north west) --++(90:33cm);

	% DESIGN II LINES
	
	\draw let \p1 = (stage1align.west), \p2=($(C-II) !.5! (R2-II)$) in node[treatment] at (\x1, \y2) (A-II) {A \\ \small{$A_{1} = 1$}};
	
	\draw let \p1 = (stage1align.west), \p2=($(F-II) !.5! (R3-II)$) in node[treatment] at (\x1, \y2) (B-II) {B \\ \small{$A_{1} = -1$}};
	
	\draw let \p1 = (designII.north), \p2=($(A-II.north) !.5! (B-II.south)$) in node[randomize] at (\x1, \y2) (R1-II) {R};
	
%		\draw let \p3 = (stage1.north) \p4=($(C-II.north) !.5! (E-II.south)$) in node[treatment] at (\x3, \y4) (test) {A};

	\draw[trialarrow] (R1-II) -- (A-II.west);
	\draw (A-II.east) -- (b1-II.center);
	\draw (A-II.east) -- (b3-II.center);
	\draw (b1-II.center) -- node [rlabel] {Responders} (b2-II.center);
	\draw (b3-II.center) -- node [nrlabel] {Non-Responders} (R2-II);
	\draw[trialarrow] (b2-II.center) -- (C-II);
	\draw[trialarrow] (R2-II) -- (D-II.west);
	\draw[trialarrow] (R2-II) -- (E-II.west);
	
	\draw[trialarrow] (R1-II) -- (B-II.west);
	\draw (B-II.east) -- (b4-II.center);
	\draw (B-II.east) -- (b6-II.center);
	\draw (b4-II.center) -- node [rlabel] {Responders} (b5-II.center);
	\draw[trialarrow] (b5-II.center) -- (F-II);
	\draw[trialarrow] (b6-II.center) -- node [nrlabel] {Non-Responders} (R3-II);
	\draw[trialarrow] (R3-II) -- (G-II.west);
	\draw[trialarrow] (R3-II) -- (H-II.west);		
		
	% TIME LINES
	
	\draw (time0) -- (time1) -- (time2);
\end{tikzpicture}

%% file: smart-design--autism.tikz
\begin{tikzpicture}[%
	node distance=6mm,
	randomize/.style={
		circle,
		minimum size=6mm,
		thick,
		black,
		draw
	},
	rerand/.style={
		xshift = 15mm
	},
	treatment/.style={
		rectangle,
		minimum size = 7mm,
		thick,
		black,
		anchor=west,
		align=center,
		draw
	},
	blank/.style={
		rectangle,
		minimum size=2mm
	},
	subgroup/.style={
		rectangle,
		rounded corners=1.5mm,
		thin,
		minimum size=6mm,
		black,
		draw
	},
	rlabel/.style={
		above,
		xshift=1mm,
		align=center
	},
	nrlabel/.style={
		below,
		xshift=-1mm,
		align=left
	},
	trialarrow/.style={
		thick,
		decoration={markings,mark=at position 1 with  {\arrow[scale=1.5,>=stealth]{>}}},
		postaction={decorate}
	},
	subsetarrow/.style={
		thin,
		decoration={markings,mark=at position 1 with {\arrow[scale=1.2,>=stealth]{>}}},
		postaction={decorate}
	}, thick]
					
	\matrix[row sep=-2mm,column sep=12mm] { %
		& \node [align=center,xshift=6mm] {\large \textbf{Stage 1}}; & & & \node [align=center,xshift=5mm] {\large \textbf{Stage 2}}; \\[5mm]
		
		% DESIGN III
		
		\node (designIII) [blank] {}; & & \node (b1-III) [blank] {}; & \node (b2-III) [blank] {}; & \node (C-III) [treatment] {C \\ $A_{2} = 0$}; \\
		& \node (stage1align) [treatment,white] {A}; & & & \\
		& & & & \node (D-III) [treatment] {D \\ $A_{2} = 1$}; \\
		& & \node (b3-III) [blank] {}; & \node (R2-III) [randomize, rerand] {R}; & \\
		& & & & \node (E-III) [treatment] {E \\ $A_{2} = -1$}; \\
		\node [randomize,white] {}; & & & &; \\	
		& & \node (b4-III) [blank] {}; & \node (b5-III) [blank] {}; & \node (F-III) [treatment] {F \\ $A_{2} = 0$}; \\
		& \node (B-III) [treatment] {B \\ $A_{1} = -1$}; & & & \\
		& & \node (b6-III) [blank] {}; & \node (b7-III) [blank] {}; & \node (G-III) [treatment] {G \\ $A_{2} = 0$}; \\[1cm]
		
		% TIMEPOINTS
		
		\node (time0) [align=center, anchor=center] {Time 0};& & & \node (time1) [align=center, anchor=center, xshift=15mm] {Time 1}; & \node (time2) [align=center, xshift = 10mm] {Time 2}; \\
	};

%		\draw[dashed,lightgray] (time1.north) -- (R2-III.south);
%		\draw[dashed,lightgray] (R2-III.north) -- (R3-II.south);
%		\draw[dashed,lightgray] (R3-II.north) --++(90:3.2cm);
%		
%		\draw[dashed,lightgray] (time2.north west) --++(90:33cm);

	% DESIGN III LINES
	
	\draw let \p1 = (stage1align.north), \p2=($(C-III) !.5! (R2-III)$) in node[treatment] at (\x1-15, \y2) (A-III) {A \\ $A_{1} = 1$};
	
	\draw let \p1 = (designIII.north), \p2=($(A-III.north) !.5! (B-III.south)$) in node[randomize] at (\x1, \y2) (R1-III) {R};
	
	\draw[trialarrow] (R1-III) -- (A-III.west);
	\draw (A-III.east) -- (b1-III.center);
	\draw (A-III.east) -- (b3-III.center);
	\draw (b1-III.center) -- node [rlabel] {Responders} (b2-III.center);
	\draw (b3-III.center) -- node [nrlabel] {Non-Responders} (R2-III);
	\draw[trialarrow] (b2-III.center) -- (C-III);
	\draw[trialarrow] (R2-III) -- (D-III.west);
	\draw[trialarrow] (R2-III) -- (E-III.west);
	
	\draw[trialarrow] (R1-III) -- (B-III.west);
	\draw (B-III.east) -- (b4-III.center);
	\draw (B-III.east) -- (b6-III.center);
	\draw (b4-III.center) -- node [rlabel] {Responders} (b5-III.center);
	\draw[trialarrow] (b5-III.center) -- (F-III);
	\draw (b6-III.center) -- node [nrlabel,align=center,xshift=1mm] {Non-Responders} (b7-III.center);
	\draw[trialarrow] (b7-III.center) -- (G-III.west);
		
	% TIME LINES
	
	\draw (time0) -- (time1) -- (time2);
\end{tikzpicture}

%% file: violS2RhoR4Delta3.tikz
% Created by tikzDevice version 0.12.2 on 2019-06-12 11:39:30
% !TEX encoding = UTF-8 Unicode
\begin{tikzpicture}[x=1pt,y=1pt]
\definecolor{fillColor}{RGB}{255,255,255}
\path[use as bounding box,fill=fillColor,fill opacity=0.00] (0,0) rectangle (289.08,260.17);
\begin{scope}
\path[clip] ( 49.20, 61.20) rectangle (263.88,210.97);
\definecolor{drawColor}{RGB}{68,1,84}

\path[draw=drawColor,line width= 0.8pt,line join=round,line cap=round] (255.93, 91.92) --
	(242.68, 92.59) --
	(229.43, 93.34) --
	(216.17, 98.04) --
	(202.92, 98.96) --
	(189.67, 99.72) --
	(176.42,108.11) --
	(163.17,110.12) --
	(149.91,113.98) --
	(136.66,125.56) --
	(123.41,131.01) --
	(110.16,140.83) --
	( 96.91,150.14) --
	( 83.65,164.99) --
	( 70.40,183.03) --
	( 57.15,201.90);
\definecolor{fillColor}{RGB}{68,1,84}

\path[fill=fillColor] (253.45, 89.44) --
	(258.40, 89.44) --
	(258.40, 94.39) --
	(253.45, 94.39) --
	cycle;

\path[fill=fillColor] (240.20, 90.11) --
	(245.15, 90.11) --
	(245.15, 95.06) --
	(240.20, 95.06) --
	cycle;

\path[fill=fillColor] (226.95, 90.87) --
	(231.90, 90.87) --
	(231.90, 95.82) --
	(226.95, 95.82) --
	cycle;

\path[fill=fillColor] (213.70, 95.56) --
	(218.65, 95.56) --
	(218.65,100.51) --
	(213.70,100.51) --
	cycle;

\path[fill=fillColor] (200.45, 96.49) --
	(205.40, 96.49) --
	(205.40,101.44) --
	(200.45,101.44) --
	cycle;

\path[fill=fillColor] (187.19, 97.24) --
	(192.14, 97.24) --
	(192.14,102.19) --
	(187.19,102.19) --
	cycle;

\path[fill=fillColor] (173.94,105.63) --
	(178.89,105.63) --
	(178.89,110.58) --
	(173.94,110.58) --
	cycle;

\path[fill=fillColor] (160.69,107.65) --
	(165.64,107.65) --
	(165.64,112.60) --
	(160.69,112.60) --
	cycle;

\path[fill=fillColor] (147.44,111.50) --
	(152.39,111.50) --
	(152.39,116.45) --
	(147.44,116.45) --
	cycle;

\path[fill=fillColor] (134.19,123.08) --
	(139.14,123.08) --
	(139.14,128.03) --
	(134.19,128.03) --
	cycle;

\path[fill=fillColor] (120.94,128.54) --
	(125.89,128.54) --
	(125.89,133.49) --
	(120.94,133.49) --
	cycle;

\path[fill=fillColor] (107.68,138.35) --
	(112.63,138.35) --
	(112.63,143.30) --
	(107.68,143.30) --
	cycle;

\path[fill=fillColor] ( 94.43,147.66) --
	( 99.38,147.66) --
	( 99.38,152.61) --
	( 94.43,152.61) --
	cycle;

\path[fill=fillColor] ( 81.18,162.51) --
	( 86.13,162.51) --
	( 86.13,167.46) --
	( 81.18,167.46) --
	cycle;

\path[fill=fillColor] ( 67.93,180.55) --
	( 72.88,180.55) --
	( 72.88,185.50) --
	( 67.93,185.50) --
	cycle;

\path[fill=fillColor] ( 54.68,199.43) --
	( 59.63,199.43) --
	( 59.63,204.38) --
	( 54.68,204.38) --
	cycle;
\end{scope}
\begin{scope}
\path[clip] (  0.00,  0.00) rectangle (289.08,260.17);
\definecolor{drawColor}{RGB}{0,0,0}

\path[draw=drawColor,line width= 0.4pt,line join=round,line cap=round] ( 57.15, 61.20) -- (216.17, 61.20);

\path[draw=drawColor,line width= 0.4pt,line join=round,line cap=round] ( 57.15, 61.20) -- ( 57.15, 55.20);

\path[draw=drawColor,line width= 0.4pt,line join=round,line cap=round] (110.16, 61.20) -- (110.16, 55.20);

\path[draw=drawColor,line width= 0.4pt,line join=round,line cap=round] (163.17, 61.20) -- (163.17, 55.20);

\path[draw=drawColor,line width= 0.4pt,line join=round,line cap=round] (216.17, 61.20) -- (216.17, 55.20);

\node[text=drawColor,anchor=base,inner sep=0pt, outer sep=0pt, scale=  1.00] at ( 57.15, 39.60) {0.0};

\node[text=drawColor,anchor=base,inner sep=0pt, outer sep=0pt, scale=  1.00] at (110.16, 39.60) {0.2};

\node[text=drawColor,anchor=base,inner sep=0pt, outer sep=0pt, scale=  1.00] at (163.17, 39.60) {0.4};

\node[text=drawColor,anchor=base,inner sep=0pt, outer sep=0pt, scale=  1.00] at (216.17, 39.60) {0.6};

\path[draw=drawColor,line width= 0.4pt,line join=round,line cap=round] ( 49.20, 67.00) -- ( 49.20,192.84);

\path[draw=drawColor,line width= 0.4pt,line join=round,line cap=round] ( 49.20, 67.00) -- ( 43.20, 67.00);

\path[draw=drawColor,line width= 0.4pt,line join=round,line cap=round] ( 49.20, 92.17) -- ( 43.20, 92.17);

\path[draw=drawColor,line width= 0.4pt,line join=round,line cap=round] ( 49.20,117.34) -- ( 43.20,117.34);

\path[draw=drawColor,line width= 0.4pt,line join=round,line cap=round] ( 49.20,142.50) -- ( 43.20,142.50);

\path[draw=drawColor,line width= 0.4pt,line join=round,line cap=round] ( 49.20,167.67) -- ( 43.20,167.67);

\path[draw=drawColor,line width= 0.4pt,line join=round,line cap=round] ( 49.20,192.84) -- ( 43.20,192.84);

\node[text=drawColor,rotate= 90.00,anchor=base,inner sep=0pt, outer sep=0pt, scale=  1.00] at ( 34.80, 67.00) {0.3};

\node[text=drawColor,rotate= 90.00,anchor=base,inner sep=0pt, outer sep=0pt, scale=  1.00] at ( 34.80, 92.17) {0.4};

\node[text=drawColor,rotate= 90.00,anchor=base,inner sep=0pt, outer sep=0pt, scale=  1.00] at ( 34.80,117.34) {0.5};

\node[text=drawColor,rotate= 90.00,anchor=base,inner sep=0pt, outer sep=0pt, scale=  1.00] at ( 34.80,142.50) {0.6};

\node[text=drawColor,rotate= 90.00,anchor=base,inner sep=0pt, outer sep=0pt, scale=  1.00] at ( 34.80,167.67) {0.7};

\node[text=drawColor,rotate= 90.00,anchor=base,inner sep=0pt, outer sep=0pt, scale=  1.00] at ( 34.80,192.84) {0.8};

\path[draw=drawColor,line width= 0.4pt,line join=round,line cap=round] ( 49.20, 61.20) --
	(263.88, 61.20) --
	(263.88,210.97) --
	( 49.20,210.97) --
	( 49.20, 61.20);
\end{scope}
\begin{scope}
\path[clip] (  0.00,  0.00) rectangle (289.08,260.17);
\definecolor{drawColor}{RGB}{0,0,0}

\node[text=drawColor,anchor=base,inner sep=0pt, outer sep=0pt, scale=  1.00] at (156.54, 15.60) {$\rho_{\mathrm{guess}} - \rho$};

\node[text=drawColor,rotate= 90.00,anchor=base,inner sep=0pt, outer sep=0pt, scale=  1.00] at ( 10.80,136.09) {Empirical power};
\end{scope}
\begin{scope}
\path[clip] ( 49.20, 61.20) rectangle (263.88,210.97);
\definecolor{drawColor}{RGB}{49,104,142}

\path[draw=drawColor,line width= 0.8pt,line join=round,line cap=round] (202.92,149.38) --
	(189.67,146.61) --
	(176.42,152.32) --
	(163.17,152.91) --
	(149.91,153.91) --
	(136.66,157.86) --
	(123.41,164.57) --
	(110.16,166.33) --
	( 96.91,179.08) --
	( 83.65,183.78) --
	( 70.40,193.68) --
	( 57.15,201.40);
\definecolor{fillColor}{RGB}{49,104,142}

\path[fill=fillColor] (202.92,149.38) circle (  2.48);

\path[fill=fillColor] (189.67,146.61) circle (  2.48);

\path[fill=fillColor] (176.42,152.32) circle (  2.48);

\path[fill=fillColor] (163.17,152.91) circle (  2.48);

\path[fill=fillColor] (149.91,153.91) circle (  2.48);

\path[fill=fillColor] (136.66,157.86) circle (  2.48);

\path[fill=fillColor] (123.41,164.57) circle (  2.48);

\path[fill=fillColor] (110.16,166.33) circle (  2.48);

\path[fill=fillColor] ( 96.91,179.08) circle (  2.48);

\path[fill=fillColor] ( 83.65,183.78) circle (  2.48);

\path[fill=fillColor] ( 70.40,193.68) circle (  2.48);

\path[fill=fillColor] ( 57.15,201.40) circle (  2.48);
\definecolor{drawColor}{RGB}{53,183,121}

\path[draw=drawColor,line width= 0.8pt,line join=round,line cap=round] (123.41,185.63) --
	(110.16,188.14) --
	( 96.91,188.73) --
	( 83.65,191.50) --
	( 70.40,192.42) --
	( 57.15,200.73);
\definecolor{fillColor}{RGB}{53,183,121}

\path[fill=fillColor] (123.41,189.47) --
	(126.74,183.70) --
	(120.08,183.70) --
	cycle;

\path[fill=fillColor] (110.16,191.99) --
	(113.49,186.22) --
	(106.83,186.22) --
	cycle;

\path[fill=fillColor] ( 96.91,192.58) --
	(100.24,186.81) --
	( 93.57,186.81) --
	cycle;

\path[fill=fillColor] ( 83.65,195.35) --
	( 86.99,189.57) --
	( 80.32,189.57) --
	cycle;

\path[fill=fillColor] ( 70.40,196.27) --
	( 73.74,190.50) --
	( 67.07,190.50) --
	cycle;

\path[fill=fillColor] ( 57.15,204.58) --
	( 60.48,198.80) --
	( 53.82,198.80) --
	cycle;
\definecolor{drawColor}{RGB}{0,0,0}

\path[draw=drawColor,line width= 0.4pt,line join=round,line cap=round] ( 49.20,104.40) rectangle (130.06, 61.20);
\definecolor{drawColor}{RGB}{53,183,121}

\path[draw=drawColor,line width= 0.4pt,line join=round,line cap=round] ( 51.63, 93.60) -- ( 67.83, 93.60);
\definecolor{drawColor}{RGB}{49,104,142}

\path[draw=drawColor,line width= 0.4pt,line join=round,line cap=round] ( 51.63, 82.80) -- ( 67.83, 82.80);
\definecolor{drawColor}{RGB}{68,1,84}

\path[draw=drawColor,line width= 0.4pt,line join=round,line cap=round] ( 51.63, 72.00) -- ( 67.83, 72.00);

\path[fill=fillColor] ( 59.73, 96.75) --
	( 62.46, 92.03) --
	( 57.00, 92.03) --
	cycle;
\definecolor{fillColor}{RGB}{49,104,142}

\path[fill=fillColor] ( 59.73, 82.80) circle (  2.02);
\definecolor{fillColor}{RGB}{68,1,84}

\path[fill=fillColor] ( 57.70, 69.97) --
	( 61.75, 69.97) --
	( 61.75, 74.03) --
	( 57.70, 74.03) --
	cycle;
\definecolor{drawColor}{RGB}{0,0,0}

\node[text=drawColor,anchor=base west,inner sep=0pt, outer sep=0pt, scale=  0.90] at ( 75.93, 90.50) {$\rho_{\mathrm{guess}} = 0.3$};

\node[text=drawColor,anchor=base west,inner sep=0pt, outer sep=0pt, scale=  0.90] at ( 75.93, 79.70) {$\rho_{\mathrm{guess}} = 0.6$};

\node[text=drawColor,anchor=base west,inner sep=0pt, outer sep=0pt, scale=  0.90] at ( 75.93, 68.90) {$\rho_{\mathrm{guess}} = 0.8$};
\end{scope}
\end{tikzpicture}